\documentclass[sigconf]{acmart}

\usepackage{tabularx}

\AtBeginDocument{%
  \providecommand\BibTeX{{%
    \normalfont B\kern-0.5em{\scshape i\kern-0.25em b}\kern-0.8em\TeX}}}

\copyrightyear{2024}
\acmYear{2024}
\setcopyright{rightsretained}
\acmConference[JCDL '24]{The 2024 ACM/IEEE Joint Conference on Digital Libraries}{December 16--20, 2024}{Hong Kong, China}
\acmBooktitle{The 2024 ACM/IEEE Joint Conference on Digital Libraries (JCDL '24), December 16--20, 2024, Hong Kong, China}\acmDOI{10.1145/3677389.3702557}
\acmISBN{979-8-4007-1093-3/24/12}

\begin{document}

\title[A Library Perspective on Supervised Text Processing in Digital Libraries]{A Library Perspective on Supervised Text Processing in Digital Libraries: An Investigation in the Biomedical Domain}

\author{Hermann Kroll}
\email{krollh@acm.org}
\orcid{0000-0001-9887-9276}
\affiliation{%
  \institution{Institute for Information Systems, \\TU Braunschweig}
  \streetaddress{Mühlenpfordtstr. 23}
  \city{Braunschweig}
  \country{Germany}
  \postcode{38106}
} 

\author{Pascal Sackhoff}
\email{p.sackhoff@tu-bs.de}
\affiliation{%
  \institution{Institute for Information Systems, \\TU Braunschweig}
  \streetaddress{Mühlenpfordtstr. 23}
  \city{Braunschweig}
  \country{Germany}
  \postcode{38106}
}   

\author{Bill Matthias Thang}
\email{m.thang@tu-bs.de}
\affiliation{%
  \institution{Institute for Information Systems, \\TU Braunschweig}
  \streetaddress{Mühlenpfordtstr. 23}
  \city{Braunschweig}
  \country{Germany}
  \postcode{38106}
}   
 
\author{Maha Ksouri}
\email{m.ksouri@tu-bs.de}
\affiliation{%
  \institution{Institute for Information Systems, \\TU Braunschweig}
  \streetaddress{Mühlenpfordtstr. 23}
  \city{Braunschweig}
  \country{Germany}
  \postcode{38106}
}

\author{Wolf-Tilo Balke}
\email{balke@ifis.cs.tu-bs.de}
\orcid{0000-0002-5443-1215}
\affiliation{%
  \institution{Institute for Information Systems, \\TU Braunschweig}
  \streetaddress{Mühlenpfordtstr. 23}
  \city{Braunschweig}
  \country{Germany}
  \postcode{38106}
}

\renewcommand{\shortauthors}{Kroll et al.}

\begin{abstract}
Digital libraries that maintain extensive textual collections may want to further enrich their content for certain downstream applications, e.g., building knowledge graphs, semantic enrichment of documents, or implementing novel access paths. All of these applications require some text processing, either to identify relevant entities, extract semantic relationships between them, or to classify documents into some categories. However, implementing reliable, supervised workflows can become quite challenging for a digital library because suitable training data must be crafted, and reliable models must be trained. While many works focus on achieving the highest accuracy on some benchmarks, we tackle the problem from a digital library practitioner. In other words, we also consider trade-offs between accuracy and application costs, dive into training data generation through distant supervision and large language models such as ChatGPT, LLama, and Olmo, and discuss how to design final pipelines. Therefore, we focus on relation extraction and text classification, using the showcase of eight biomedical benchmarks. 

\end{abstract}

\begin{CCSXML}
<ccs2012>
   <concept>
       <concept_id>10002951.10003317.10003347.10003352</concept_id>
       <concept_desc>Information systems~Information extraction</concept_desc>
       <concept_significance>500</concept_significance>
       </concept>
   <concept>
       <concept_id>10002951.10003260.10003277.10003279</concept_id>
       <concept_desc>Information systems~Data extraction and integration</concept_desc>
       <concept_significance>300</concept_significance>
       </concept>
 </ccs2012>
\end{CCSXML}

\ccsdesc[500]{Information systems~Information extraction}
\ccsdesc[300]{Information systems~Data extraction and integration}

\keywords{Text Processing, Supervision, Digital Libraries}

\maketitle

\section{Introduction}
One way to explore a digital library's content is to apply natural language processing methods, e.g., identify central entities (e.g., the Person Albert Einstein), their relationships (e.g., Albert Einstein was born in Ulm), and classify documents as belonging to classes (e.g., descriptive articles).
The extraction of semantic relationships between named entities is already used in several digital library projects for different purposes, e.g., constructing a biomedical knowledge graph from scientific papers like SemMedDB~\cite{DBLP:journals/bioinformatics/KilicogluSFRR12}, harvesting leader boards of how computer science methods perform on benchmarks~\cite{DBLP:journals/jodl/KabongoDA24}, harvesting scientific information as done in SciGraph~\cite{DBLP:conf/eeke/YanC22}, enabling graph-based discovery systems in digital libraries~\cite{DBLP:journals/jodl/KrollPKKRB24}, or enriching library content like newspapers as done in the Swiss-Luxembourgish impresso~\cite{impresso}.

While we know that information extraction workflows can be beneficial for a digital library, as the previous projects have successfully demonstrated, the costs of designing such workflows are typically high.
One reason is that these workflows usually require supervision. 
That is, system designers must know what they are looking for in advance, and several examples must be given to train supervised extraction methods.
While alternatives like unsupervised extraction workflows~\cite{DBLP:conf/jcdl/KrollPPB22a,DBLP:journals/jodl/KrollPPB24} might be seen as a remedy here because they bypass the need for training data in the extraction step completely, they usually require extensive filtering and may not achieve the best quality. 
So, supervision is typically required when implementing high-quality, reliable workflows that result in a canonicalized knowledge representation.

From a natural language processing perspective, several works exist that propose advanced methods for extracting named entities and their semantic relationships or classifying texts in general; see~\cite{DBLP:journals/ftdb/WeikumDRS21,DBLP:journals/iswa/DetrojaBB23,DBLP:journals/csur/SmirnovaC19} to name just a few.
When implementing extraction workflows in a digital library, questions beyond a benchmark-centric evaluation arise, e.g., about trade-offs between costs and quality. 
Regarding training and application costs, a cheaper model might be favored over a complex model, achieving higher accuracy. 
In brief, this work is written from the perspective of a digital library. 
It differs from existing work in that we 1) compare the trade-off between extraction quality and costs, 2) dive into designing complete end-to-end systems in contrast to benchmark-centric evaluations, and 3) approach how we can generate/retrieve training data. 

Our investigation is focused on the biomedical domain as a showcase and centers around two example tasks: relation extraction between named entities and text classification. 
For each task, we experimented with four different benchmarks for generalizability. 
For reproducibility, we share our code and data at our GitHub repository\footnote{\url{https://github.com/HermannKroll/SupervisedTextProcessing}},\footnote{Software Heritage ID: \href{https://archive.softwareheritage.org/swh:1:dir:047fbde2d12b0fd4a12f8fdc4ab5347a6744f893}{swh:1:dir:047fbde2d12b0fd4a12f8fdc4ab5347a6744f893}}.
We tackle three research questions in this work:

\textbf{RQ1}: \textit{Which model should we use in a digital library project? Can we still rely on older models (e.g., SVMs and Random Forests), or must we use the latest language models? If so, which one? The largest one? A domain-specific one? Or a generic one? In brief, we investigate the trade-off between quality and application costs. }

\textbf{RQ2}: \textit{Another, often neglected, question is how to design a full digital library pipeline. Should we train a single model that is capable of doing multi-tasks simultaneously (e.g., extracting relationships between drugs, diseases, and genes), or should we train models for each task separately, e.g., one model to predict relationships between drugs and diseases and one model for drugs and genes?}

\textbf{RQ3}: \textit{How can we label training data? Do we still require experts to annotate the data, or can we rely on methods like weak supervision? Or can we completely rely on large language models like ChatGPT4o and LLama 3 that annotate our data? What happens if our data includes noise and thus may only have a moderate quality?}

\section{Related Work}
\paragraph{Named Entity Recognition and Disambiguation.}
The first step in extracting semantic relationships between named entities is to identify these entities in the text.
Usually, recognition tools \textit{recognize} entities within texts, and subsequent disambiguation tools assign those text spans to precise identifiers to disambiguate them.
A comprehensive overview of possible detection methods is given in~\cite{DBLP:journals/ftdb/WeikumDRS21}.
A plethora of different tools exist to identify biomedical entities in texts, e.g., PubTator~\cite{10.1093/nar/gkae235}, GNormPlus~\cite{gnormplus}, GNorm2~\cite{DBLP:journals/bioinformatics/WeiLILL23}, TaggerOne~\cite{DBLP:journals/bioinformatics/LeamanL16}, and many more. 
While entity detection is a relevant topic in digital libraries, our work focuses on relation extraction between them and thus assumes that the entities are given. 

\paragraph{Relation Extraction.}
Relation extraction is a well-studied task in natural language processing; see~\cite{DBLP:journals/iswa/DetrojaBB23} for an overview of methods, ~\cite{DBLP:journals/csur/SmirnovaC19} for a survey on distantly supervised methods, and~\cite{DBLP:journals/ftdb/WeikumDRS21} for general strategies to create and curate knowledge bases.

Milosevic and Thielemann compared different biomedical relationship extraction methods and models for their applicability on different benchmarks~\cite{DBLP:journals/ws/MilosevicT23}. 
One of their findings was that BERT-based models showed the highest performance, even when compared to larger models like T5.
Another finding involved high expenses for training data generation. 
%\enquote{\textit{On the market, the pricing of a single annotated sentence can range between 1–3 euros, depending on the complexity of the task [...] The commissioned manual annotations of our data set (around 7000 sentences in total) cost 16,200 euros. The further cost comes from cloud infrastructure and machine learning engineering. Costs in developing relationship extraction models and approaches remain among the main challenges.}}
Another study focused on whether we should use model tuning or prompt tuning for the relation extraction task~\cite{DBLP:journals/jbi/PengYSYCBW24}. 
Lai et al.~\cite{DBLP:journals/jbi/LaiWLCL23} trained a single model, called BioRex, on several biomedical relation extraction benchmarks. 
They, therefore, integrated different benchmarks involving sentence-wide, document-level-wide, and n-ary relation extraction. 
They showed that their model achieved a high F1 score across all benchmarks. 
Instead, our work is placed as a discussion and investigation of how a digital library should approach supervised text processing.

Weakly supervised methods like Snorkel~\cite{DBLP:journals/vldb/RatnerBEFWR20} have been introduced to automatically generate training data by using some noisy labeling functions, e.g., based on hand-crafted rules or by using existing knowledge bases. 
An application of Snorkel in the biomedical domain can be found here~\cite{DBLP:journals/cj/KumarS24}.
Methods like Snorkel reduce the costs of labeling data, and the idea is that if enough data is labeled even in a noisy fashion, the models will have a \textit{similar} quality compared to models that are trained on high-quality but less training data (labeled by experts).
We will apply a similar strategy to create noisy labels in our investigation.

Generating training data for LLMs has already been studied recently; see~\cite{DBLP:conf/acl/ChiaBPS22,DBLP:conf/emnlp/JosifoskiSP023,DBLP:conf/emnlp/LiZL023}.
Josifoski et al.~\cite{DBLP:conf/emnlp/JosifoskiSP023} generated training data by LLMs and used it to train another model, called SynthIE, for the information extraction task. 
The authors demonstrated that this model performed well on existing benchmarks but also noted that generating training data via LLMs can introduce bias. 
Chia et al.~\cite{DBLP:conf/acl/ChiaBPS22} introduced RelationPrompt and showed that language models can effectively generate synthetic training data for unseen relations.
Li et al.~\cite{DBLP:conf/emnlp/LiZL023} summarized the advantages and limitations of generating training data by LLMs for the text classification task.
They found that models trained on synthetic data usually have a decreased performance on tasks with higher levels of subjectivity.
More works focus on generating training data by LLMs for even more tasks, e.g., for the retrieval task~\cite{DBLP:conf/naacl/ThakurNAWLC24}.
In contrast to these works, our work explores the usage of LLMs for training data generation in the biomedical domain and from a practitioner's perspective.

Our study differs from existing ones by going beyond comparing precision, recall, and F1. 
We also focus on:
1) different strategies to generate training data with weak supervision and LLMs.
2) trade-offs between accuracy vs. training and application costs, which is especially relevant for digital libraries but rarely considered.

\paragraph{Text Classification.}
An overview of text classification methods can be found in~\cite{DBLP:journals/tist/LiPLXYSYH22}. 
Again, our focus is on something other than inventing a new text classification method in our paper. 
Instead, we investigate which methods can be applied, how they differ in accuracy and costs, and what happens if we introduce noise into the training sets, i.e., how stable/robust different models are. 

\section{Experimental Setup and Used Models}
In the following, we briefly describe our setup and environment.

\subsection{Models and Vectorization}
\label{sec:models}
For our investigation, we compared traditional classification models to language models.
We selected the following traditional models:
1) \textbf{SVC (Support Vector Classifier)}:
An SVC provides high accuracy in classification tasks, effectively handles complex decision boundaries and high-dimensional data, and, uses kernel functions to capture non-linear relationships in the data.
2) \textbf{XGBoost (Extreme Gradient Boosting Classifier)}:
XGBoost is known for its high performance and efficiency in handling large datasets and missing data.
It utilizes regularization and gradient boosting, which sequentially adds models to correct errors of previous models, leading to improved accuracy.
3) \textbf{Random Forest Classifier}:
A Random Forest offers robustness against overfitting due to ensemble learning techniques.
It is effective in capturing complex relationships by building multiple decision trees.

These traditional classification models require the transformation of texts into a vector representation. 
Therefore, we used two different strategies: tf-idf and sentence transformers. 
1) tf-idf is a traditional vectorization technique based on the frequency of words in the document and across the corpus. It is a common method but may lack the ability to capture semantic similarities and context, especially in complex sentence structures.
2) Sentence transformers allow us to derive a semantic representation of a text by utilizing pre-trained transformer models like BERT to generate fixed-length vector representations of sentences~\cite{DBLP:conf/emnlp/ReimersG19}.
They capture semantic meaning and context by considering the entire sentence rather than individual words, which is ideal for tasks where understanding the context and meaning of sentences is crucial.
Thus, they may perform better than tf-idf, especially in capturing semantic similarities and nuances.
We used the sBERT implementation with the \textit{all-MiniLM-L6-v2} model.

Language Models do not require the transformation of texts into a vector representation a-priori, as they come with built-in encoders/decoders. For our investigation, we used three generic models: \textbf{BERT}~\cite{devlin2019bert} (\href{https://huggingface.co/google-bert/bert-base-uncased}{bert-base-uncased}),  \textbf{RoBERTa}~\cite{liu2019roberta} (\href{http://www.overleaf.com}{roberta-base}),  and \textbf{XLNet}~\cite{DBLP:conf/nips/YangDYCSL19} (\href{https://huggingface.co/xlnet/xlnet-base-cased}{xlnet-base-cased}).
We compared the generic models (not domain-specific) against three biomedical language models that have been pre-trained on biomedical texts:
\textbf{BioBERT}~\cite{lee2020biobert} (\href{https://huggingface.co/dmis-lab/biobert-v1.1}{dmis-lab/biobert-v1.1}), \textbf{BioLinkBERT}~\cite{DBLP:conf/acl/YasunagaLL22} (\href{https://huggingface.co/michiyasunaga/BioLinkBERT-base}{michiyasunaga/Bio\\LinkBERT-base}), and \textbf{PubMedBERT}~\cite{DBLP:journals/health/GuTCLULNGP22} (\href{https://huggingface.co/microsoft/BiomedNLP-BiomedBERT-base-uncased-abstract}{microsoft/BiomedNLP-PubMedBERT-base-uncased-abstract}).

\begin{table}
    \centering
    \begin{tabularx}{0.48\textwidth}{l|X}
         Model & Hyperparameter Grid \\
         \toprule
         SVC & C: \{0.1, 1, 10, 100\}, kernel: \{poly, rbf, sigmoid\}, degree: \{1, 2, 3, 4, 5, 6\}\\
         \midrule
         XGBoost & n\_estimators: \{50, 100\}, max\_depth: \{3, 5\}, learning\_rate: \{0.01, 0.1\}, subsample: \{0.8, 1.0\}, colsample\_bytree: \{0.8, 1.0\}\\
         \midrule
         Random Forest & n\_estimators: \{50, 100\}, max\_depth: \{None, 10, 20\}, min\_samples\_split': \{2, 5\}, min\_samples\_leaf: \{1, 2\}\\
         \midrule
         Language Models & learning\_rate: \{1e-3, 1e-4, 1e-5\}, epochs: \{1, 3, 5\}, weight\_decay: \{0.0, 0.1, 0.2, 0.3\}\\
    \end{tabularx}
    \caption{Set of tested hyperparameters for each model.}
    \label{tab:models_hs}
\end{table}

\paragraph{Hyperparameter Tuning: GridSearch}
Training a model also means selecting some training parameters like the kernal type or a lerning rate. 
However, testing different parameter combinations is usually beneficial and may strongly affect the final classification accuracy. 
That is why we used a grid search to find optimal parameters, i.e., a model is trained on a train data set, parameters are optimized on a development set, and the final model is evaluated on a test set.
The set of tested parameters for each model is listed in Table~\ref{tab:models_hs}.

\subsection{Hardware and Environment}
We performed the experiments on our server, which has two Intel(R) Xeon(R) Gold 6336Y CPU @ 2.40GHz (24 cores and 48 threads each), 2TB DDR4 main memory, and nine Nvidia A40 GPUs with 48GB memory. 
We limited the number of parallel workers on the CPU to 32, as we observed a decrease when using too many parallel workers due to I/O and communication overhead. 
Moreover, we used only a single GPU to train/apply language models and provide meaningful runtime measurements. 
Our implementation is written in Python. 
We used the default sklearn and hugging face implementation to train our classification models. For details, see our repository.

\section{Task 1: Relation Extraction}
Our first investigation focuses on the relation extraction task, i.e., extracting semantic relationships between named entities. 

\subsection{Data Sets}
We used the following benchmarks: SemEval-2013 Task 9~\cite{DDI:DBLP:journals/jbi/Herrero-ZazoSMD13} containing drug-drug interactions (DDI), BioCreative V Track 3~\cite{CDR:DBLP:journals/biodb/WeiPLDMLWL16}, containing chemical-disease relations (CDR), and BioCreative VI Track 5~\cite{ChemProt:krallinger2017overview}, containing chemical-protein interactions (ChemProt).

\begin{table}[t]
    \centering
    \resizebox{.45\textwidth}{!}{
    \begin{tabular}{l|rrr|rrr|rrr|rrr}
        Model & \multicolumn{3}{c}{CDR}  & \multicolumn{3}{c}{ChemProtC}  & \multicolumn{3}{c}{ChemProtE}  & \multicolumn{3}{c}{DDI}  \\
        & P & R & F1 & P & R & F1& P & R & F1& P & R & F1 \\
        \toprule

        \multicolumn{13}{c}{Traditional Classification Models} \\
        \midrule
        SVC + tfidf & \textbf{0.49} & 0.58 & \textbf{0.53} & \textbf{0.38} & \textbf{0.75} & \textbf{0.51} & \textbf{0.46} & \textbf{0.59} & \textbf{0.46} & \textbf{0.22} & 0.81 & \textbf{0.35}  \\
        SVC + sBERT & \textbf{0.49} & 0.58 & \textbf{0.53} & \textbf{0.38} & \textbf{0.75} & \textbf{0.51} & \textbf{0.46} & \textbf{0.59} & \textbf{0.46} & \textbf{0.22} & 0.81 & \textbf{0.35}  \\
        XGBoost + tfidf & 0.45 & 0.63 & \textbf{0.53} & 0.36 & 0.57 & 0.44 & 0.45 & 0.53 & 0.45 & 0.21 & 0.78 & 0.34  \\
        XGBoost + sBERT & 0.45 & 0.63 & \textbf{0.53} & 0.36 & 0.57 & 0.44 & 0.45 & 0.53 & 0.45 & 0.21 & 0.78 & 0.34  \\
        Random Forrest + tfidf & 0.39 & 0.61 & 0.47 & 0.32 & 0.72 & 0.44 & 0.42 & 0.48 & 0.42 & 0.18 & 0.92 & 0.3  \\
        Random Forrest + sBERT & 0.43 & \textbf{0.69} & \textbf{0.53} & 0.32 & 0.72 & 0.44 & 0.42 & 0.49 & 0.42 & 0.18 & \textbf{0.93} & 0.3  \\
        \midrule
        \multicolumn{13}{c}{Language Models} \\
        \midrule
        BERT & 0.57 & 0.7 & 0.63 & 0.56 & 0.74 & 0.59 & 0.47 & 0.83 & 0.6 & 0.56 & 0.93 & 0.7  \\
        RoBERTa & 0.57 & 0.75 & 0.65 & 0.56 & 0.74 & 0.58 & 0.52 & 0.77 & 0.62 & 0.54 & 0.93 & 0.68  \\
        XLNet & 0.54 & 0.55 & 0.55 & 0.56 & 0.74 & 0.58 & 0.48 & 0.81 & 0.6 & 0.59 & 0.88 & 0.71  \\
        BioLinkBERT & \textbf{0.59} & \textbf{0.79} & \textbf{0.68} & 0.62 & \textbf{0.82} & \textbf{0.67} & 0.57 & 0.86 & \textbf{0.69} & \textbf{0.67} & 0.92 & \textbf{0.78}  \\
        BioBERT & 0.58 & 0.8 & \textbf{0.68} & 0.6 & 0.81 & 0.64 & 0.56 & \textbf{0.87} & 0.68 & 0.59 & 0.92 & 0.72  \\
        PubMedBERT & 0.6 & 0.78 & \textbf{0.68} & \textbf{0.63} & 0.81 & \textbf{0.67} & \textbf{0.58} & 0.86 & \textbf{0.69} & 0.59 & \textbf{0.94} & 0.73  \\
    \end{tabular}
    }
    \caption{Task 1 (RE). We report the relation extraction quality (precision, recall, F1) when comparing several models on the test data sets of the corresponding benchmarks.}
    \label{tab:re_quality}
\end{table}

\textbf{Drug-Drug-Interaction (DDI)~\cite{DDI:DBLP:journals/jbi/Herrero-ZazoSMD13}}:
The DDI benchmark is divided into two subtasks from which we only used the relation extraction part. 
The task is to find sentences containing two or more drugs that influence each other in the body, i.e., the effect of one drug is altered by the other or both cause some side effect.
Only training and test data sets are provided.
We split the train data in half to create a development set to optimize hyperparameters.

\textbf{Chemical-Disease-Relation (CDR)~\cite{CDR:DBLP:journals/biodb/WeiPLDMLWL16}}:
We utilized the benchmark's relation extraction subtask, asking specifically for chemical-induced disease relations, i.e., a chemical induces a disease if it is responsible for the appearance of the disease.

\textbf{Chemical-Protein-Interaction (ChemProt)~\cite{ChemProt:krallinger2017overview}}:
The dataset is a pure relation extraction benchmark with annotated proteins, genes, and chemicals.
It asks for regulations between chemicals and proteins.
In contrast to the other benchmarks, which require a binary classification, the ChemProt data is labeled into ten different relation types, of which five are considered in the benchmark evaluation (as denoted by the benchmark authors).
A problem with that benchmark is that for some relations, less than 174 examples are given (CPR:5—Agonist), which makes it difficult to train reliable models. 
We followed ideas of related work like~\cite{DBLP:journals/jbi/LaiWLCL23} that used that benchmark but grouped some specific relations into more general ones.
We grouped the relations agonist-inhibitor, antagonist, indirect-downregulator, and inhibitor into down-regulation.
The relations activator, agonist, agonist-activator, and indirect-upregulator are grouped into up-regulation. 
Every other relation is grouped to no regulation.
We called this benchmark ChemProtC (C for complex), asking for up, down, and no regulations. 
We also created an alternative version, called ChemProtE (E for easy), by grouping the relations up/down regulations and substrates together.
This benchmark asks whether a chemical regulates a protein.

\begin{table}[t]
    \centering
    \begin{tabular}{l|ccc}
        Benchmark & \#Train & \#Dev & \#Test \\
        \midrule
        CDR~\cite{CDR:DBLP:journals/biodb/WeiPLDMLWL16}& 500/4662& 500/4651& 500/4853\\
        ChemProt~\cite{ChemProt:krallinger2017overview}& 1020/10311& 612/6243& 800/8140\\
        DDI~\cite{DDI:DBLP:journals/jbi/Herrero-ZazoSMD13}& 714/6976& -/-& 191/1299\\
        
    \end{tabular}
    \caption{Task 1 (RE). Benchmark doc./sent. distribution.}
    \label{tab:re_datasets}
\end{table}

\textbf{Setup.}
The number of documents and training sentences is listed in Table~\ref{tab:re_datasets}. 
For the DDI benchmark, the sentences were already given together with every possible drug-drug relation combination and the correct label.
Both the CDR and ChemProt datasets came with full abstracts, which had to be processed first.
Therefore, we used the spaCy sentenciser to split the abstracts into sentences. 
Next, if the benchmark states that a certain relation is expressed in a document between entities a and b, we consider all sentences that contain a and b as training examples.
Sentences that contain other entities are considered as negative examples.
That is why the number of labeled sentences (in combination with an entity pair) in the datasets is much higher than the number of raw sentences.

For the training, hyper-parameter search, and evaluation, we used an obfuscation technique to ensure that the models tried to learn the sentence structure instead of the entity names if they occurred more frequently.
For that, we replaced the entities of a given relation with the strings \textit{<entity1>} and \textit{<entity2>}, e.g., \textit{"Tomudex (ZD1694) is a specific antifolate-based thymidylate synthase inhibitor active in a variety of solid tumor malignancies."} was changed to  \textit{"<entity1> (ZD1694) is a specific antifolate-based <entity2> inhibitor active in a variety of solid tumor malignancies."}

\begin{figure}
    \centering
    \includegraphics[width=0.90\linewidth]{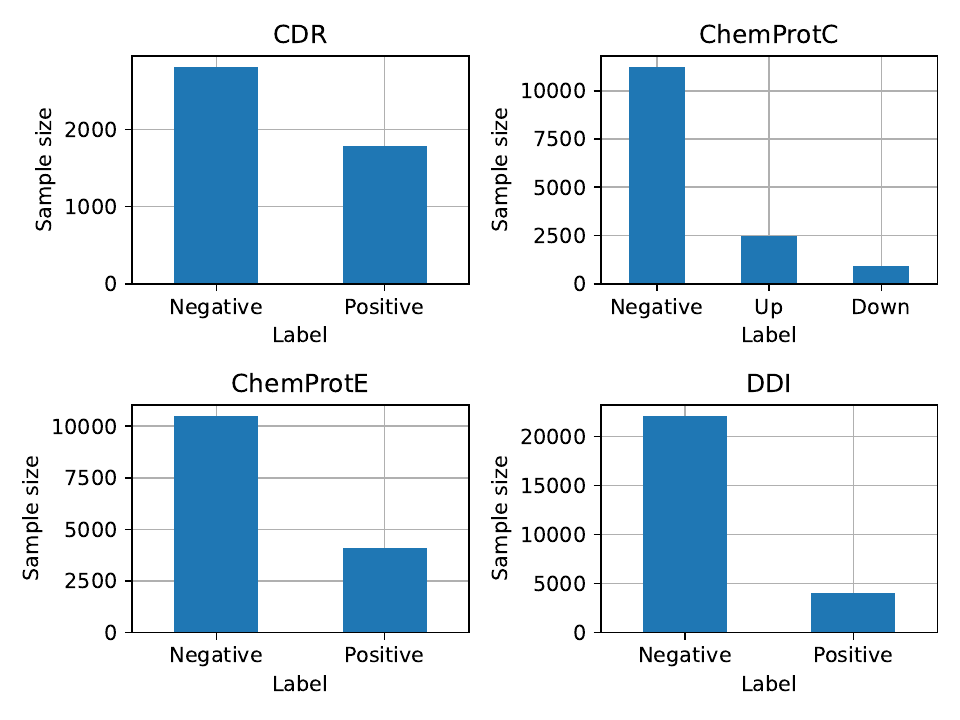}
    \caption{Task 1 (RE). Label distribution of each benchmark.}
    \Description{Task 1 (RE). Label distribution of each benchmark.}
    \label{fig:re_label_distribution}
\end{figure}
The label distribution for each benchmark is listed in Figure~\ref{fig:re_label_distribution}.
First, the negative class is dominant across all benchmarks. 
In brief, CDR is the best-balanced dataset (2810 negative and 1785 positive examples).
For the remaining benchmarks, the negative class is very dominant, which is why we used downsampling for training.

\begin{table}[t]
    \centering
    \resizebox{.45\textwidth}{!}{
    \begin{tabular}{l|rrrr}
        Model & Training  & HS & Application & ET PubMed \\
        \toprule

        \multicolumn{5}{c}{Traditional Classification Models} \\
        \midrule
        SVC + tfidf           & 30 s & 4 min 13 s & 9.79e-04 s & 4 d 0 h 51 min 34.51 s \\
        SVC + sBERT           & 31 s & 4 min 40 s & 9.70e-04 s & 3 d 23 h 54 min 19.56 s \\
        XGBoost + tfidf       & \textbf{2 s} & 3 min 54 s & \textbf{5.63e-05 s} & \textbf{5 h 33 min 56.68 s} \\
        XGBoost + sBERT       & \textbf{2 s} & 3 min 54 s & 7.00e-05 s & 6 h 55 min 27.09 \\
        Random Forest + tfidf & 11 s & \textbf{37 s} & 7.96e-05 s & 7 h 52 min 5.52 s \\
        Random Forest + sBERT & 11 s & 39 s & 7.68e-05 s & 7 h 35 min 55.17 s \\
        \midrule
        \multicolumn{5}{c}{Language Models on CPU} \\
        \midrule
        BERT        & 1 h 0 min 9 s & 32 h 56 min 16 s & 2.74e-02 s & 112 d 18 h 40 min 20.45 s\\
        RoBERTa     & 1 h 7 min 7 s & 23 h 57 min 6 s & 2.34e-02 s & 96 d 12 h 10 min 48.29 s\\
        XLNet       & 41 min 8 s & 60 h 45 min 32 s & 5.24e-02 s & 216 d 0 h 6 min 58.48 s\\
        BioBERT     & 23 min 5 s & 38 h 2 min 8 s & 3.09e-02 s & 127 d 12 h 27 min 28.23 s\\
        BioLinkBERT & 5 min 58 s & 10 h 50 min & 2.36e-02 s & 97 d 1 h 22 min 46.96 s\\
        PubMedBERT  & \textbf{5 min 47 s} & \textbf{10 h 43 min 21 s} & \textbf{2.31e-02 s} & \textbf{95 d 6 h 30 min 3.06 s}\\
        \midrule
        \multicolumn{5}{c}{Language Models on GPU} \\
        \midrule
        BERT & 2 min 17 s & 1 h 22 min 15 s & 3.67e-03 s & 15 d 2 h 34 min 48.94 s \\
        RoBERTa & 3 min 53 s & 1 h 36 min 59 s & 3.50e-03 s & 14 d 10 h 28 min 32.65 s \\
        XLNet & 25 min 51 s & 6 h 56 min 31 s & 1.10e-02 s & 45 d 5 h 1 min 54.96 s \\
        BioBERT & 50 s & 1 h 30 min 11 s & 3.64e-03 s & 14 d 23 h 51 min 42.48 s \\
        BioLinkBERT & 1 min 28 s & \textbf{52 min 32 s} & \textbf{3.38e-03 s} & \textbf{13 d 22 h 14 min 3.62 s} \\
        PubMedBERT & \textbf{32 s} & 53 min 53 s & \textbf{3.38e-03 s} & 13 d 22 h 37 min 51.36 s \\
    \end{tabular}
    }
    \caption{Task 1 (RE). Runtimes were measured on DDI (as it was the largest benchmark). The training and HS time is reported in total, whereas the application time is normalized by the number of sentences in the test set. ET PubMed is the estimated time when applying the model to 37M documents. }
    \label{tab:re_time}
\end{table}

\subsection{RQ1: Relation Extraction}
We then trained our models on the train sets and optimized hyperparameters (see Table~\ref{tab:models_hs}) via a grid search on validation. 
Before training, we balanced the train sets using randomized downsampling, ensuring that the number of samples in each class matched that of the smallest class. 
This downsampling process was made reproducible by setting a specific random seed, allowing for consistent and repeatable results across different runs.
We report the final classification results on test sets.
The results are listed in Table~\ref{tab:re_quality}.

As expected, the language models result in better predictions than the shallow models regarding the F1-score.
The differences between the shallow and language models (LM) having the maximum F1-scores are between 0.15 for CDR (0.53 - shallow, 0.68 - LM) and 0.43 points for DDI (0.35 - shallow, 0.78 - LM).
In terms of the F1 scores, the SVC was the best shallow model, followed by the XGBoost models.
BioLinkBERT and PubMedBERT achieved the best F1 scores across all benchmarks for the language models.
We saw that the language models offered the highest scores when classifying semantic relationships between entities, which we expected.

\paragraph{Hyperparameter Search} 
Our next question was focused on the influence of the hyperparameter selection. 
So, how many configurations should be tested, and what can a digital library then expect? Is it worth it, and how many configurations need to be explored?

\begin{figure}
    \centering
    \includegraphics[width=0.95\linewidth]{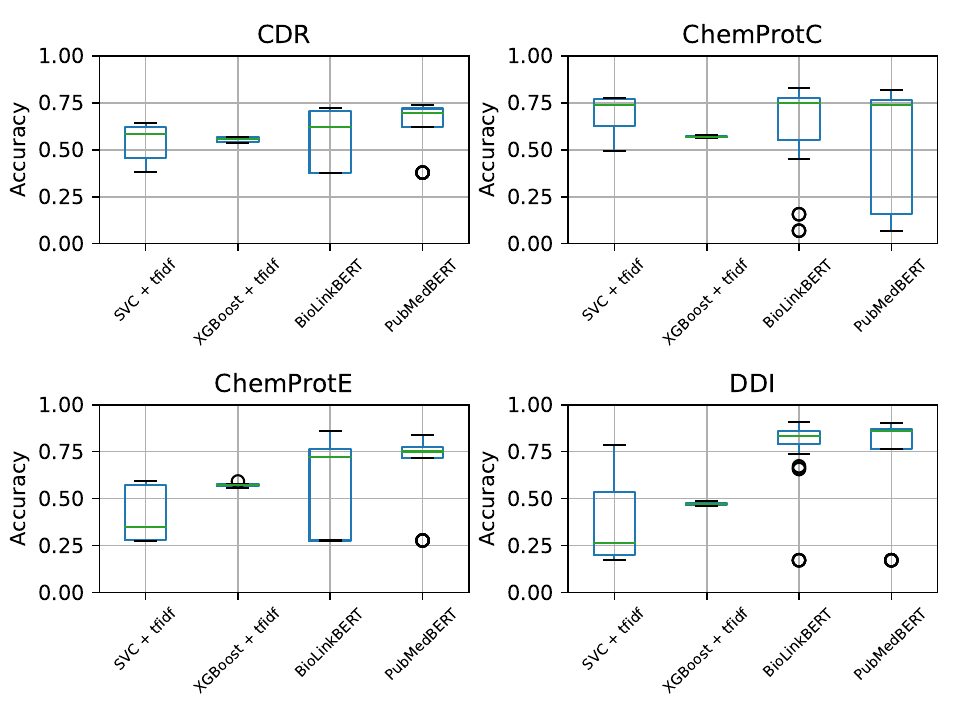}
    \caption{Task 1 (RE). Hyperparameter search distribution between the best and worst models for the two best shallow models and language models comparing the accuracy score.}
    \Description{Task 1 (RE). Hyperparameter search distribution between the best and worst models for the two best shallow models and language models comparing the accuracy score.}
    \label{fig:re_hs_comparison}
\end{figure}

To visualize the impact of hyperparameter selection, we present box plots in Figure~\ref{fig:re_hs_comparison} for the top-performing models across the four datasets. Each box plot shows the range of accuracy scores obtained using a certain hyperparameter combination for training and testing the trained model on a test set.
The results indicate that the LMs generally performed more robustly across various hyperparameter settings. However, even these models, such as BioLinkBERT, showed some sensitivity to hyperparameter choices, evident in the range of accuracy scores on CDR. 
SVC showed greater variability with lower median accuracy, suggesting a greater sensitivity to hyperparameter choices. In contrast, the XGBoost model was very robust to hyperparameter tuning.
Notably, the ChemProtC dataset reveals some anomalies, likely due to our re-grouping of the data. This was particularly evident with the PubMedBERT model, which had a wider range of performance scores, including some very low values.
Overall, our findings suggest that investing in extensive hyperparameter search can be beneficial.

\paragraph{Application Costs}
We measured the training, hyperparameter search (HS), and application times of the best-performing models on DDI. 
We selected DDI because it was the largest benchmark. The results are listed in Table~\ref{tab:re_time}.
For CPU application time, we evaluated the test set as a single batch. On GPU, we upsampled the combined train, dev, and test sets to 1M balanced samples to also measure the memory transfer overhead between main memory and GPU memory, processing batches of 256 for precise runtime estimation. Times were normalized by sample count for comparability and allowing us to estimate the time required to process the entire PubMed collection ($\sim$37M documents, as of July 2024). 
The estimated time (ET) is calculated by multiplying the \textit{time per sentence} with the \textit{number of documents on PubMed} (36,555,430) with the \textit{average sentences per document} (9.74).
The average sentences per document is calculated by dividing the sum of all documents (4837) of the used benchmarks divided by the sum of all sentences (47135) retrieved with the SpaCy sentenciser.
Note that this estimation is somewhat an overestimation because it assumes that all documents have a similar number of sentences and that each sentence contains entities and are thus relevant to predict a relation.

Table~\ref{tab:re_time} shows that language models (LMs) generally required longer training and HS times compared to traditional models. However, applying LMs, especially on GPUs, was significantly faster. 
Among LMs, XLNet was notably slower across all stages than BERT-based models. This extended time could be attributed to XLNet's more complex architecture and different fine-tuning processes, which involved segment-level recurrence and a bidirectional context, leading to higher computational demands.
In summary, while traditional models offered quick training and tuning, language models, especially PubMedBERT, provided a better performance with acceptable application times. Traditional models are ideal for tasks where inference speed is crucial and GPUs are not available. The choice of a model thus depends on an application's requirements.

\begin{table}
    \centering
    \resizebox{.45\textwidth}{!}{
    \begin{tabular}{l|rrr|rrr|rrr|rrr}
        Model &  \multicolumn{3}{c}{CDR}  & \multicolumn{3}{c}{ChemProtC}  & \multicolumn{3}{c}{ChemProtE}  & \multicolumn{3}{c}{DDI}  \\
       &  P & R & F1 & P & R & F1& P & R & F1& P & R & F1 \\
        \toprule
        \multicolumn{13}{c}{SingleTask Learning} \\
        \midrule
        SVC + tfidf & 0.49 & 0.58 & 0.53 & 0.38 & 0.75 & 0.51 & 0.46 & 0.59 & 0.46 & 0.22 & 0.81 & 0.35 \\
        XGBoost + tfidf & 0.45 & 0.63 & 0.53 & 0.36 & 0.57 & 0.44 & 0.45 & 0.53 & 0.45 & 0.21 & 0.78 & 0.34 \\
        BioLinkBERT & 0.59 & \textbf{0.79} & \textbf{0.68} & 0.62 & \textbf{0.82} & \textbf{0.67} & 0.57 & \textbf{0.86} & \textbf{0.69} & \textbf{0.67} & 0.92 & \textbf{0.78} \\
        PubMedBERT & \textbf{0.6} & 0.78 & \textbf{0.68} & \textbf{0.63} & 0.81 & \textbf{0.67} & \textbf{0.58} & \textbf{0.86} & \textbf{0.69} & 0.59 & \textbf{0.94} & 0.73 \\

        \midrule
        \multicolumn{13}{c}{MultiTask Learning} \\
        \midrule
        SVC + tfidf & 0.24 & 0.19 & 0.18 & 0.26 & 0.31 & 0.21 & - & - & - & 0.22 & 0.2 & 0.1  \\
        XGBoost + tfidf & 0.03 & 0.0 & 0.0 & 0.03 & 0.0 & 0.0 & - & - & - & 0.02 & 0.0 & 0.0  \\
        
        BioLinkBERT & \textbf{0.29} & \textbf{0.27} & \textbf{0.25} & \textbf{0.33} & \textbf{0.46} & 0.32 & - & - & - & 0.27 & \textbf{0.32} & \textbf{0.27}  \\
        PubMedBERT & 0.28 & 0.26 & 0.24 & \textbf{0.33} & \textbf{0.46} & \textbf{0.33} & - & - & - & \textbf{0.28} & \textbf{0.32} & \textbf{0.27}  \\

    \end{tabular}
    }
    \caption{Task 1 (RE). SingleTask vs. MultiTask learning evaluated on the test sets of the biomedical benchmarks.} 
    \label{tab:re_stvsmt}
\end{table}

\subsection{RQ2: System Architecture}
Our second research question asked how to design an entire system pipeline in a digital library. 
Basically, two setups are possible: 

\textbf{1) SingleTask Learning.} We could train a model that predicts relationships between certain entity-type combinations.
Here, one model for drug-disease relationships, one for drug-drug interactions, and one for chemical-protein interactions. 
Deploying such a model in a digital library then requires a pre-selection step that works as follows:
First, detect entities within a sentence. Compute the set of all entity pairs in that sentence. Check the entities' types for each pair and select the corresponding prediction model.

\textbf{2) MultiTask Learning.} We could train a single model that learns to predict all possible relations by combining the training data from different benchmarks into a single training set. 
This model should then learn to extract every known relation in one step.
We combined DDI, ChemProtC, and CDR training data for our setup. 
We ignored ChemPromtE here because it contains the same training examples (but labeled differently), which would cause confusion within the model. The same sentence with the same entities requires classifying an up-regulation as well as just a regulation.

The results of both setups are listed in Table~\ref{tab:re_stvsmt}.
We compared four models: the top two shallow architectures and the top two language models, ranked by their classification quality.
The MultiTask strategy results reveal problems in every of the evaluated benchmarks.
This might come due to the rather low-size downsampling of the training data because of the ChemProtC \textit{Downregulator} subclass (with 936 samples).
As expected, the LMs work better than the shallow models.
Surprisingly, the multiclass ChemProtC task works the best, whereas the CDR task results are more than half as good as those of the SingleTask strategy.
In brief, the findings support that SingleTask learning performs better here (or that more sophisticated strategies for MultiTask learning are required). 

\subsection{RQ3: Data Labeling}
Our last research question concerns how digital libraries can/should label their training data when relying on supervised classification models. 
The advantages of asking experts are rather obvious: The data will likely be of good quality, but costs are high, processes may take time, and resulting data is usually limited.

\textit{Distantly-Supervised Labeling.}
The related work section describes weak supervision as a possible remedy~\cite{DBLP:journals/vldb/RatnerBEFWR20}. 
The central idea is that external knowledge is used to label sentences. 
If some sentence includes two entities and the entities have a relationship within the given knowledge base, then we implicitly assume that the sentence also expresses this relationship. 
In brief, distant supervision allows fast and large data set generation.
That is why we investigate it here. 
However, it requires external knowledge bases that include the relations someone is interested in, and it might also be limited in precision, as sentences may be labeled in a noisy fashion.

% DrugBank Database from 	2024-03-14, DrugBank Vocabulary 	2024-03-14 https://go.drugbank.com/releases/latest
We used simple string comparisons to check whether a relation between two entities exists in such a knowledge base.
Additionally, the entities were expanded with their corresponding synonyms to increase the chance of finding relations based on the different writing styles of the entities.
Because CDR contains additional Medical Subject Headings (MeSH) information for each entity an additional comparison based on the MeSH IDs was conducted.

The \textit{Comparative Toxigenomics Database}~\cite{CTD:DBLP:journals/nar/DavisWJSWM23} knowledge base was used for the CDR and ChemProt data set. 
Specifically, the \textit{Chemical-disease associations} together with the \textit{chemical}- and \textit{disease}- vocabularies for CDR and the \textit{Chemical-gene interactions} together with the \textit{chemical}- and \textit{gene}-vocabularies.
The selection already reveals some problems: 
A chemical-disease association does not directly imply that a chemical induces a disease. 
Hence, the labeling will be very noisy by design.
For DDI, we used the latest release (14-03-2024) of the \textit{DrugBank Database}~\cite{Drugbank:DBLP:journals/nar/WishartKGSHSCW06} together with the \textit{DrugBank vocabulary}.
Please note that we did not find a suitable knowledge base that expresses up and down regulations between chemicals and proteins, so we could not use distant supervision to relabel the ChemProtC data set, so we skipped the experiment.

\textit{Large Language Models (LLMs).}
LLMs like ChatGPT are trained on large-scale data via an instruction-driven style. 
In other words, we may assume that they have comprehensive real-world knowledge contained inside them and that they understand task instructions, e.g., labeling a sentence based on some criterion.
We expect that the generation of training data via large language models is thus less cost-intensive than asking human experts but likely more cost-intensive than distant supervision due to high computational costs (when executing the model on local hardware) or high costs when using external APIs.
So, can we use LLMs to label our data?

Consider the DDI benchmark. 
The benchmark asks to predict whether a sentence describes an interaction between two drugs. 
For each of our benchmarks, we created three different prompts that follow a basic structure but differ in their description of an interaction. 
One advantage is that the benchmark only asks whether an interaction is described.
In other words, we can prompt the language model to only answer with \textit{yes} or \textit{no}.

Together with a pharmaceutical domain expert (PhD, years of experience in pharmaceutical research), we created the following prompts (\{0\} - sentence, \{1\} - drug 1, \{2\} - drug 2):

\textbf{Prompt 1.} \textit{Consider the following sentence: \{0\}. Does this sentence describes the information that \{1\} interacts with \{2\}. Interacts describes that \{1\} interacts with \{2\}, e.g., as a drug-drug interaction, via a shared target or via some mechanism. Answer only with yes or no.}

\textbf{Prompt 2.} \textit{Consider the following sentence: \{0\}. Does this sentence describes the information that \{1\} interacts with \{2\}. Interacts means that \{1\} has a reaction with \{2\}. Answer only with yes or no.}

Similarly, we built prompts for CDR (chemical induces some disease) and ChemProtE (chemical regulates some protein). 
For ChemProtC, which asks for up/down/no regulations, we created prompts like the following:

\textbf{Chem.-Prot. Prompt 1.} \textit{Consider the following sentence: \{0\}. Does this sentence describe the information that \{1\} regulates \{2\}? We ask for three kinds of regulations between \{1\} and \{2\}: (1) up regulation means that \{1\} is an activator, agonist, agonist-activator or indirect-upregulator of \{2\}. (2) down regulation means that \{1\} is an agonist-inhibitor, antagonist, indirect-downregulator or inhibitor of \{2\}. (3) no means that there is no regulation. Answer only with up, down, or no.}

The other prompts can be found at our repository.
Please note that we did not perform any fine-tuning of the prompts. 
They were created in a single two-hour session with one domain expert. 
Our goal was to answer our research question by just building intuitive prompts.
Of course, someone could argue that tuning the prompts would result in higher results, e.g., using few-shot prompting.
While we agree, the central problem is that this labeling strategy should be applied when NO training data is available. 
Training prompts then are unclear and would require human expert labeling again.

We compared the following LLMs: OlMo 7B~\cite{olmo} (allenai/OLMo-7B-Instruct, open data, open model, open source), Llama 3 8B~\cite{llama3} (meta-llama/Meta-Llama-3-8B-Instruct, open model, open source and closed data, requires a usage verification), the biomedical LLM BioMistral 7B~\cite{DBLP:conf/acl/LabrakBMGRD24} (BioMistral/BioMistral-7B, open model, trained for biomedical purposes, open source and closed data) and ChatGPT-4o~\cite{gpt4o} (everything closed). 
We downloaded the first three models and performed the experiments.
For GPT-4o~\cite{gpt4o}, we used the official API and spent about 130\$ to relabel all four benchmarks.
Note that we only labeled the training sets. 
We tested four different strategies to assign the final labels: 
1) Use the first prompt and decide based solely on that prompt (1P).
Use all three prompts for a sentence and perform a vote over the results.
2) 1-Yes is sufficient. (3P-1Y)
3) 2-Yes are sufficient. (3P-2Y)
4) 3-Yes are sufficient. (3P-3Y)
For 2) and 3), we implemented early stopping, i.e., if the first prompts are answered with yes (depending on whether one or two yes are required), the remaining prompt(s) is not shown to reduce the runtime.
In the case of ChemProtC, the model must deliver the same answer (up/down/no).
We counted cases where an LLM replied with an answer that did not correspond to one answer option.

\begin{table*}[t]
    \centering
    \resizebox{.95\textwidth}{!}{
    \begin{tabular}{l|r|rrrr|rrrr|rrrr|rrrr}
        && \multicolumn{4}{c}{CDR}  & \multicolumn{4}{c}{ChemProtC}  & \multicolumn{4}{c}{ChemProtE}  & \multicolumn{4}{c}{DDI} \\
        Training Data Generation & $t_{sen}$$\downarrow$ & P & R & F1 & NA & P & R & F1 & NA & P & R & F1 & NA & P & R & F1 & NA \\
        \toprule
        Distantly-supervised  & $<0.1$s & 0.44 & \textbf{0.99} & 0.61 & - & - & - & - &- & 0.43 & 0.05 & 0.08 & - & 0.14 & 0.27 & 0.19 & -\\
        \midrule
     
        OlMo 7B (1-Prompt)        & 0.15s & 0.52 & 0.95 & 0.67 & 1 & 0.14 & 0.01 & 0.02 & 7768 & 0.36 & 0.77 & 0.49 & 11 & 0.18 & 0.98 & 0.30 & 2 \\
        OlMo 7B (3-Prompt, 1-Yes) & 0.26s & 0.52 & 0.96 & 0.67 & - & 0.15 & 0.04 & 0.07 & 5273 & 0.35 & \textbf{0.89} & 0.50 &  5 & 0.17 & \textbf{0.99} & 0.29 & - \\
        OlMo 7B (3-Prompt, 2-Yes) & 0.38s & 0.60 & 0.89 & 0.71 & - & 0.14 & 0.02 & 0.03 & 5273 & 0.36 & 0.77 & 0.49 &  5 & 0.18 & 0.98 & 0.31 & - \\
        OlMo 7B (3-Prompt, 3-Yes) & 0.45s & 0.64 & 0.83 & \textbf{0.72} & - & 0.12 & 0.01 & 0.01 & 5273 & 0.37 & 0.61 & 0.46 &  5 & 0.22 & 0.94 & 0.35 & - \\

        \midrule

        Llama 3 8B (1-Prompt)        & 0.2s  & 0.74 & 0.42 & 0.54 & - & 0.05 & 0.20 & 0.08 & 1503 & 0.41 & 0.56 & 0.48 & - & 0.26 & 0.81 & 0.39 & - \\
        Llama 3 8B (3-Prompt, 1-Yes) & 0.37s & 0.74 & 0.65 & 0.69 & - & 0.04 & \textbf{0.62} & 0.08 &  246 & 0.41 & 0.70 & 0.52 & - & 0.24 & 0.91 & 0.38 & - \\
        Llama 3 8B (3-Prompt, 2-Yes) & 0.51s & 0.76 & 0.46 & 0.57 & - & 0.05 & 0.52 & 0.09 &  246 & 0.45 & 0.56 & 0.50 & - & 0.26 & 0.84 & 0.40 & - \\
        Llama 3 8B (3-Prompt, 3-Yes) & 0.59s & 0.73 & 0.26 & 0.39 & - & 0.05 & 0.29 & 0.08 &  246 & 0.50 & 0.43 & 0.47 & - & 0.28 & 0.72 & 0.41 & - \\
        \midrule
     
        BioMistral 7B (1-Prompt)        & 0.17s & 0.55 & 0.44 & 0.49 & 418 & 0.18 & 0.02 & 0.03 & 7985 & 0.34 & 0.23 & 0.27 & 588 & 0.19 & 0.37 & 0.25 & 2278 \\
        BioMistral 7B (3-Prompt, 1-Yes) & 0.43s & 0.51 & 0.76 & 0.61 &   4 & 0.17 & 0.08 & \textbf{0.11} & 2426 & 0.32 & 0.61 & 0.42 &   1 & 0.17 & 0.91 & 0.29 &   19 \\
        BioMistral 7B (3-Prompt, 2-Yes) & 0.54s & 0.66 & 0.41 & 0.50 &   4 & \textbf{0.21} & 0.01 & 0.01 & 2426 & 0.37 & 0.25 & 0.30 &   1 & 0.21 & 0.64 & 0.31 &   19 \\
        BioMistral 7B (3-Prompt, 3-Yes) & 0.56s & 0.71 & 0.11 & 0.19 &   4 & 0.0  & 0.0  & 0.0  & 2426 & 0.42 & 0.06 & 0.10 &   1 & 0.25 & 0.24 & 0.24 &   19 \\
        \midrule

        GPT-4o (1-Prompt)        & - & 0.77 & 0.59 & 0.67 & - & 0.03 & 0.04 & 0.03 & 2 & 0.50 & 0.55 & 0.52 & - & 0.55 & 0.88 & 0.67 & - \\
        GPT-4o (3-Prompt, 1-Yes) & - & 0.75 & 0.67 & 0.71 & - & 0.03 & 0.05 & 0.03 & - & 0.49 & 0.79 & \textbf{0.60} & - & 0.53 & 0.93 & \textbf{0.68} & - \\
        GPT-4o (3-Prompt, 2-Yes) & - & 0.77 & 0.59 & 0.67 & - & 0.03 & 0.03 & 0.03 & - & 0.54 & 0.57 & 0.56 & - & 0.55 & 0.89 & \textbf{0.68} & - \\
        GPT-4o (3-Prompt, 3-Yes) & - & \textbf{0.78} & 0.50 & 0.61 & - & 0.03 & 0.02 & 0.02 & - & \textbf{0.60} & 0.48 & 0.53 & - & \textbf{0.56} & 0.84 & 0.67 & - \\

    \end{tabular}
    }
    \caption{Task 1 (RE). We report the relabeling quality (precision, recall, F1, not answered) of distant supervision and prompting LLMs on each benchmark's training set. The time is normalized to labeling one  sentence (averaged over all benchmarks).}
    \label{tab:re_labeling}
\end{table*}

\paragraph{Data Labeling Results.}
The results of our data relabeling are shown in Table~\ref{tab:re_labeling}.
The Table shows the precision (P), recall (R), F1-score (F1), and not answered (NA) based on the original expert labeling of the training sets of each benchmark. 
Note, that NA is the sum of results of unrelated answers such as \textit{the} for ChemProtC instead of the expected classes (up, down, no) which counted as false negative.
Focusing on distant supervision (distantly supervised), the strategy resulted in a high recall and an acceptable precision for CDR whereas the other benchmarks (ChemProtE, DDI) resulted in low-quality results. 
There are noticeable differences between the LLMs in the quality of answers.
OlMo produced more than adequate results but with many NAs (7768@1P  5273@any-3Y) at ChemProtC.
For ChemProtE and DDI, OlMo had the best recall (but low precision) with the 3P-1Y strategy and, on the other hand, good precision on average at CDR.
The results of the LLama model were mostly more precise at a lower recall (CDR, ChemProtE, DDI).
LLama had the lowest error on ChemProtC, resulting in a high recall (0.62) but extremely low precision (0.04) compared to the other free-to-use models.
BioMistral, as the domain-specific model, however, had difficulties answering any of the strategies without errors, and, except for ChemProtC, the results were worse than those of other LLMs at any benchmark. 
In other words, predicting yes and no is much more reliable than asking for certain relations.
This could be caused by the lack of prompt optimization or the possibility that the chosen LLMs cannot answer multiple-choice questions for this specific task.
GPT-4o resulted in the best results regarding precision in the binary benchmarks but had issues at ChemProtC.
Surprisingly, the model answered with a not matching response in only two cases (ChemProtC, 1P).
Concerning the runtimes, the LLMs labeled a sentence between 0.15 and 0.59s, which is reasonable.
The mean time to answer the 1P strategy was always the shortest because only one prompt was evaluated.
Looking at the recall, the 1P, the 3P-1Y, and the 3P-2Y strategies had the best results.
For precision, the 3P-3Y led to the best results.

\begin{table}
    \centering
    \resizebox{.45\textwidth}{!}{
    \begin{tabular}{ll|rrr|rrr|rrr|rrr}
        Model & Labeling & \multicolumn{3}{c}{CDR}  & \multicolumn{3}{c}{ChemProtC}  & \multicolumn{3}{c}{ChemProtE}  & \multicolumn{3}{c}{DDI}  \\
       & & P & R & F1 & P & R & F1& P & R & F1& P & R & F1 \\
        \toprule
        SVC + tfidf & Experts & 0.49 & 0.58 & 0.53 & 0.38 & 0.75 & 0.51 & 0.46 & 0.59 & 0.46 & 0.22 & 0.81 & 0.35 \\
        XGBoost + tfidf & Experts & 0.45 & 0.63 & 0.53 & 0.36 & 0.57 & 0.44 & 0.45 & 0.53 & 0.45 & 0.21 & 0.78 & 0.34 \\
        BioLinkBERT & Experts & 0.59 & \textbf{0.79} & \textbf{0.68} & 0.62 & \textbf{0.82} & \textbf{0.67} & 0.57 & \textbf{0.86} & \textbf{0.69} & \textbf{0.67} & 0.92 & \textbf{0.78} \\
        PubMedBERT  & Experts & \textbf{0.6} & 0.78 & \textbf{0.68} & \textbf{0.63} & 0.81 & \textbf{0.67} & \textbf{0.58} & \textbf{0.86} & \textbf{0.69} & 0.59 & \textbf{0.94} & 0.73 \\
        
        \midrule
        
        SVC + tfidf & Distant. & 0.39 & \textbf{0.79} & \textbf{0.53} & - & - & - & 0.29 & 0.03 & 0.05 & \textbf{0.21} & 0.22 & 0.22 \\
        XGBoost + tfidf & Distant. & 0.37 & 0.59 & 0.46 & - & - & - & 0.29 & 0.18 & 0.22 & 0.16 & 0.2 & 0.18  \\
        BioLinkBERT & Distant. & \textbf{0.41} & 0.78 & \textbf{0.53} & - & - & - & \textbf{0.34} & \textbf{0.43} & \textbf{0.38} & 0.17 & \textbf{0.39} & \textbf{0.23} \\
        PubMedBERT & Distant.  & \textbf{0.41} & 0.73 & \textbf{0.53} & - & - & - & 0.33 & 0.4 & 0.36 & 0.17 & \textbf{0.39} & \textbf{0.23} \\

        \midrule

        SVC + tfidf & LLama 3 (3Y) & 0.47 & 0.27 & 0.34 & 0.25 & \textbf{0.18} & \textbf{0.16} & 0.3 & \textbf{0.54} & 0.39 & 0.27 & 0.46 & 0.34  \\
        XGBoost + tfidf & LLama 3 (3Y) & 0.44 & 0.31 & 0.36 & 0.24 & 0.17 & 0.14 & 0.27 & 0.45 & 0.34 & 0.23 & 0.5 & 0.31  \\
        BioLinkBERT & LLama 3 (3Y) & 0.51 & 0.49 & 0.5 & \textbf{0.26} & 0.16 & 0.14 & \textbf{0.42} & 0.52 & \textbf{0.46} & 0.34 & 0.77 & 0.47  \\
        PubMedBERT & LLama 3 (3Y)  & \textbf{0.53} & \textbf{0.5} & \textbf{0.52} & 0.25 & 0.16 & 0.14 & \textbf{0.42} & 0.5 & 0.45 & \textbf{0.37} & \textbf{0.83} & \textbf{0.51}  \\

        \midrule

        SVC + tfidf & GPT-4o (2Y) & 0.5 & 0.39 & 0.43 & \textbf{0.33} & \textbf{0.29} & \textbf{0.28} & 0.32 & 0.57 & 0.41 & 0.24 & 0.73 & 0.36  \\
        XGBoost + tfidf & GPT-4o (2Y) & 0.46 & 0.36 & 0.41 & 0.34 & 0.32 & 0.3 & 0.3 & 0.51 & 0.37 & 0.27 & 0.72 & 0.39  \\
        BioLinkBERT & GPT-4o (2Y) & 0.55 & \textbf{0.67} & \textbf{0.61} & 0.32 & 0.22 & 0.25 & 0.42 & \textbf{0.64} & \textbf{0.5} & \textbf{0.48} & \textbf{0.95} & \textbf{0.64}  \\
        PubMedBERT & GPT-4o (2Y)  & \textbf{0.62} & 0.56 & 0.59 & 0.31 & 0.25 & 0.27 & \textbf{0.44} & 0.49 & 0.46 & \textbf{0.48} & \textbf{0.95} & \textbf{0.64}  \\

    \end{tabular}
    }
    \caption{Task 1 (RE). We trained the classification models on the noisy training data generated by distant supervision or LLMs. LLama used the \textit{3-Yes} and GPT-4o the \textit{2-Yes} strategy. We tested the trained models on the original test sets. }
    \label{tab:re_transfer}
\end{table}

\paragraph{Data Labeling Transfer Learning.}
The last experiment was designed to show whether the noisy labeling via LLMs is sufficient to train models.
While the quality decreased compared to expert labeling, the question was whether this would have a strong effect in the end.
While the training data might get noisy, the models could still minimize the problem in their learning phase.
That is why we trained our models on the noisily generated training data and tested them on the original test set.
We used the best-performing classification models with their best hyperparameter settings (we did not conduct a new search here). 
The results are shown in Table~\ref{tab:re_transfer}.
For CDR, the different labeling methods seem to perform quite well. 
The distantly supervised labeling method achieved a similar, but slightly decreased, F1 score compared to the expert labeling. This suggests that the \textit{Comparative Toxigenomics Database}~\cite{CTD:DBLP:journals/nar/DavisWJSWM23} knowledge base used for distant labeling is reliable and provides high-quality data for this task. 
On ChemProtC, the difference between the labeling methods and expert labeling was more noticeable. 
The models' performance dropped significantly when trained on noisily generated data, which could also be a cause of our re-grouping of the data.
Overall, GPT-4o mode performed the best across most tasks and labeling methods. 
It consistently outperformed other models with training on BERT models, demonstrating its robustness and effectiveness even when trained on noisily labeled data. 
This highlights the potential of advanced language models to handle noisy data and achieve high performance without the need for perfect labeling.
In brief, LLMs labeled the training data sufficiently well for our purposes and came with acceptable costs in the end.

\section{Task 2: Text Classification}
The text classification task can be formulated as follows: Given some document, classify whether its text belongs to some class. 

\subsection{Benchmarks}
We used four biomedical text classification benchmarks that provide train, dev, and test data containing abstracts and their assigned classes. 
For instance, one abstract may be classified as describing information about Long COVID or as being about Pharmaceutical Technology.
The first three benchmarks are established benchmarks.
The fourth benchmark is a benchmark that we crafted. 

\textbf{Hallmarks of cancer.} This dataset\footnote{\url{https://autonlp.ai/datasets/hoc-(hallmarks-of-cancer)}, last accessed: 07.2024} was created by Simon Baker~\cite{DBLP:journals/bioinformatics/BakerSGAHSK16, DBLP:conf/coling/BakerKP16}. He collected 1852 biomedical publication abstracts using terms for the ten cancer hallmarks. This was further refined by expanding the original ten Hallmarks of Cancer into a more detailed taxonomy with 37 classes. Annotations were made on sentence and abstract levels for texts showing clear evidence of association with one or several hallmarks~\cite{DBLP:journals/bioinformatics/BakerASPGHSK17}.

\textbf{Ohsumed.} The Ohsumed dataset\footnote{\url{https://disi.unitn.it/moschitti/corpora.htm}, last accessed: 07.2024}, created by William Hersh and his colleagues~\cite{DBLP:conf/sigir/HershBLH94}, was formulated by extracting clinical paper abstracts from the MEDLINE database, with a focus on 23 Medical Subject Headings (MeSH) disease categories. This process yielded 13,929 unique abstracts~\cite{DBLP:conf/aaai/YaoM019}. A "positive abstract" is defined as one that aligns with one of the 23 disease categories through its content and indexing terms, while a "negative abstract" is identified as one that either falls outside these categories or sufficiently relates to the designated disease categories~\cite{DBLP:conf/ecml/Joachims98}.

\textbf{Long COVID.} The Long COVID dataset was developed for classifying documents related to the long-term effects of COVID-19. It was compiled through manual curation and searches across PubMed and other databases like LitCovid. This collection provides a targeted benchmark for understanding the long-term impacts of COVID-19 in scientific literature.~\cite{DBLP:journals/biodb/LangnickelDHDF22}

\textbf{Pharmaceutical Technology (our own).} The objective of this dataset is to classify whether an article belongs to "Pharmaceutical Technology". We used a list of journals  (see our repository) that publish articles about Pharmaceutical Technology to derive document abstracts as positive training examples. This mapping was facilitated by utilizing the MEDLINE database, which contains a list of all PubMed IDs and their corresponding journal information. Articles from these journals were labeled as belonging to Pharmaceutical Technology, while negative examples were randomly sampled from other journals so that both classes were balanced.

\begin{table}[t]
    \centering
    \begin{tabular}{l|ccc}
        Benchmark & \#Train & \#Dev & \#Test \\
        \midrule
        Hallmarks of cancer & 12119 & 1798 & 3547 \\
        Ohsumed  & 5028 & 1258 & 7643 \\
        Long COVID  & 331 & 83 & 138 \\
        Pharm. Tech. & 14000 & 3000 & 3000 \\
        
    \end{tabular}
    \caption{Task 2 (TC). Benchmark document distribution.}
    \label{tab:tc_datasets}
\end{table}

\begin{figure}
    \centering
    \includegraphics[width=0.95\linewidth]{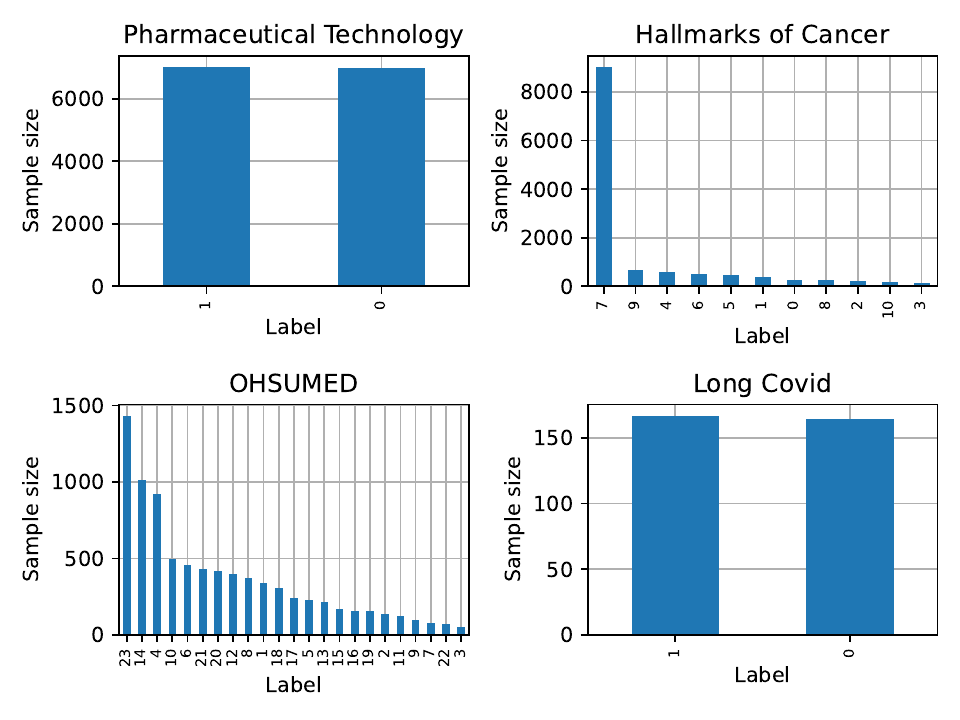}
    \caption{Task 2 (TC). Label distribution for each benchmark.}
    \Description{Task 2 (TC). Label distribution for each benchmark.}
    \label{fig:text_classificaiton_labels}
\end{figure}

The data set size is shown in Table~\ref{tab:tc_datasets}.
The assigned label distribution is shown in Figure~\ref{fig:text_classificaiton_labels}. 
The Long COVID and Pharmaceutical Technology benchmarks had an equal distribution of labels, whereas the OHSUMED and Hallmarks of Cancer benchmarks had an uneven distribution.
We again down-sampled the training data to the less frequent class to obtain a balanced training objective. 
Again, we trained our models on train, optimized hyperparameters (as listed in Table~\ref{tab:models_hs}) on development and tested our models on test.

\begin{table}[t]
    \centering
    \resizebox{.45\textwidth}{!}{
    \begin{tabular}{l|rrr|rrr|rrr|rrr}
        Model & \multicolumn{3}{c}{Hallmark}  & \multicolumn{3}{c}{Ohsumed}  & \multicolumn{3}{c}{Long COVID}  & \multicolumn{3}{c}{Pharm. Tech.}  \\
        & P & R & F1 & P & R & F1& P & R & F1& P & R & F1 \\
        \toprule

        \multicolumn{13}{c}{Traditional Classification Models} \\
        \midrule
        SVC + tfidf & \textbf{0.36} & \textbf{0.67} & \textbf{0.44} & \textbf{0.31} & \textbf{0.39} & \textbf{0.33} & \textbf{0.92} & 0.90 & \textbf{0.91} & \textbf{0.87} & \textbf{0.89} & \textbf{0.88}  \\
        SVC + sBERT & \textbf{0.36} & \textbf{0.67} & \textbf{0.44} & \textbf{0.31} & \textbf{0.39} & \textbf{0.33} & \textbf{0.92} & 0.90 & \textbf{0.91} & \textbf{0.87} & \textbf{0.89} & \textbf{0.88}  \\
        XGBoost + tfidf & 0.27 & 0.57 & 0.31 & 0.25 & 0.29 & 0.24 & 0.88 & \textbf{0.94} & \textbf{0.91} & 0.86 & 0.84 & 0.85  \\
        XGBoost + sBERT & 0.27 & 0.57 & 0.31 & 0.25 & 0.29 & 0.24 & 0.88 & \textbf{0.94} & \textbf{0.91} & 0.86 & 0.84 & 0.85  \\
        Random Forrest + tfidf & 0.24 & 0.5 & 0.26 & 0.2 & 0.24 & 0.2 & 0.89 & \textbf{0.94} & \textbf{0.91} & 0.77 & 0.86 & 0.81  \\
        Random Forrest + sBERT & 0.23 & 0.5 & 0.25 & 0.2 & 0.24 & 0.2 & 0.82 & 0.91 & 0.86 & 0.77 & 0.88 & 0.82  \\
        \midrule
        \multicolumn{13}{c}{Language Models} \\
        \midrule
        BERT & 0.36 & 0.76 & 0.44 & 0.24 & 0.33 & 0.22 & 0.92 & 0.90 & \textbf{0.91} & 0.89 & 0.90 & 0.89  \\
        RoBERTa & 0.37 & 0.75 & 0.45 & 0.23 & 0.29 & 0.21 & 0.92 & 0.88 & 0.90 & 0.88 & \textbf{0.93} & 0.91  \\
        XLNet & 0.27 & 0.64 & 0.31 & 0.33 & 0.45 & 0.36 & \textbf{0.94} & 0.85 & 0.89 & 0.90 & 0.90 & 0.90  \\
        BioBERT & 0.44 & 0.8 & 0.54 & \textbf{0.38} & 0.51 & \textbf{0.41} & 0.90 & 0.88 & 0.89 &\textbf{ 0.91} & \textbf{0.93} & \textbf{0.92}  \\
        BioLinkBERT & 0.41 & 0.8 & 0.51 & \textbf{0.38} & \textbf{0.52} & \textbf{0.41} & 0.91 & \textbf{0.91} & \textbf{0.91} & 0.90 & 0.91 & 0.91  \\
        PubMedBERT & \textbf{0.46} & \textbf{0.82} & \textbf{0.56} & \textbf{0.38} & 0.51 & \textbf{0.41} & 0.87 & \textbf{0.91} & 0.89 & 0.90 & 0.92 & 0.91  \\

    \end{tabular}
    }
    \caption{Task 2 (TC). We report the text classification quality (precision, recall, F1) when comparing several models on the test data of the corresponding benchmarks.}
    \label{tab:tc_quality}
\end{table}

%\begin{table}[t]
%    \centering
%    \resizebox{.45\textwidth}{!}{
%    \begin{tabular}{l|rrr|rrr}
%        Model & \multicolumn{3}{c}{Hallmark} & \multicolumn{3}{c}{Pharm. Tech.}  \\
%        & P & R & F1 & P & R & F1 \\
%        \toprule
%
%        \multicolumn{13}{c}{Traditional Classification Models} \\
%        \midrule
%        SVC + tfidf & \textbf{0.36} & \textbf{0.67} & \textbf{0.44} & \textbf{0.87} & \textbf{0.89} & \textbf{0.88}  \\
%        SVC + sBERT & \textbf{0.36} & \textbf{0.67} & \textbf{0.44} & \textbf{0.87} & \textbf{0.89} & \textbf{0.88}  \\
%        XGBoost + tfidf & 0.27 & 0.57 & 0.31 & 0.86 & 0.84 & 0.85  \\
%        XGBoost + sBERT & 0.27 & 0.57 & 0.31 & 0.86 & 0.84 & 0.85  \\
%        Random Forrest + tfidf & 0.24 & 0.5 & 0.26 & 0.77 & 0.86 & 0.81  \\
%        Random Forrest + sBERT & 0.23 & 0.5 & 0.25 & 0.77 & 0.88 & 0.82  \\
%        \midrule
%        \multicolumn{13}{c}{Language Models} \\
%        \midrule
%        BERT & 0.36 & 0.76 & 0.44 & 0.89 & 0.90 & 0.89  \\
%        RoBERTa & 0.37 & 0.75 & 0.45 & 0.88 & \textbf{0.93} & 0.91  \\
%        XLNet & 0.27 & 0.64 & 0.31 & 0.90 & 0.90 & 0.90  \\
%        BioBERT & 0.44 & 0.8 & 0.54 &\textbf{ 0.91} & \textbf{0.93} & \textbf{0.92}  \\
%        BioLinkBERT & 0.41 & 0.8 & 0.51 & 0.90 & 0.91 & 0.91  \\
%        PubMedBERT & \textbf{0.46} & \textbf{0.82} & \textbf{0.56} & 0.90 & 0.92 & 0.91  \\
%
%    \end{tabular}
%    }
%    \label{tab:tc_quality2}
%\end{table}

\subsection{RQ1: Text Classification}
Table~\ref{tab:tc_quality} shows the classification quality of different models across the four benchmarks described above. Among traditional models, SVC with both tfidf and sBERT configurations performed best, achieving F1 scores of 0.44 for Hallmarks of Cancer and 0.88 for Pharmaceutical Technology. XGBoost models also did well, especially in the Long COVID benchmark, with an F1 score of 0.91, while Random Forest models performed lower overall. Language models (LMs) outperformed traditional ones, with BioBERT achieving an F1 score of 0.41 in Ohsumed and 0.92 in Pharmaceutical Technology. PubMedBERT exceled with the highest F1 score of 0.56 in Hallmarks of Cancer and consistently high scores in other benchmarks, such as 0.91 in Pharmaceutical Technology.

\begin{figure}
    \centering
    \includegraphics[width=0.95\linewidth]{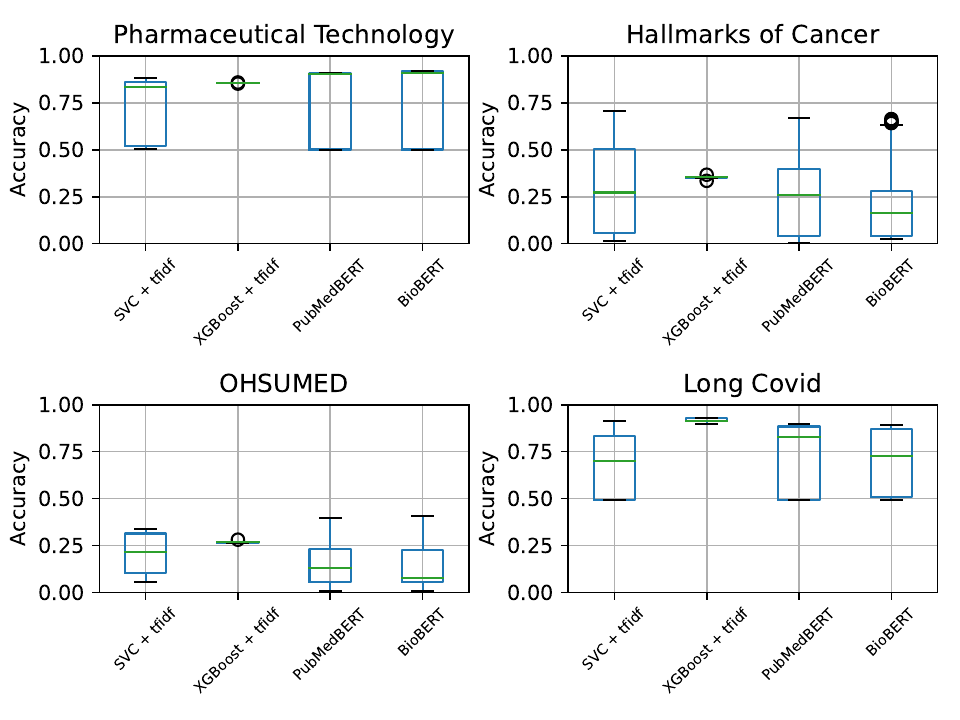}
    \caption{(TC). Hyperparameter search sistribution between the best and worst models for the two best shallow models and language models comparing the accuracy score.}
    \Description{(TC). Hyperparameter search sistribution between the best and worst models for the two best shallow models and language models comparing the accuracy score.}
    \label{fig:tc_hs_comparison}
\end{figure}

\paragraph{Hyperparameter Search}
The influence of hyperparameters is again shown in Figure~\ref{fig:tc_hs_comparison} as box plots:
The x-axis contains the best-performing models (two for each category) from Table~\ref{tab:tc_quality}. 
The y-axis shows the accuracy score based on some selected hyperparameter combinations. BioBERT showed significant sensitivity to hyperparameter choices, which is reflected in its wide range of accuracy scores across the benchmarks. PubMedBERT delivered strong performance with noticeable variability, especially in the Hallmarks of Cancer benchmark, also indicating its sensitivity to hyperparameter settings. On the other hand, SVC + tfidf showed greater variability in performance in OHSUMED and Hallmarks of Cancer benchmarks, achieving higher median accuracy, which highlights the impact of effective hyperparameter tuning. Like in the previous experiment, XGBoost showed robustness to hyperparameter tuning, achieving higher median accuracy in most cases, suggesting it is less sensitive to changes in hyperparameter settings. 
In brief, our findings indicate that hyperparameter search significantly impacts model performance, especially for LMs.

\begin{figure*}
    \centering
    \includegraphics[width=1.0\textwidth]{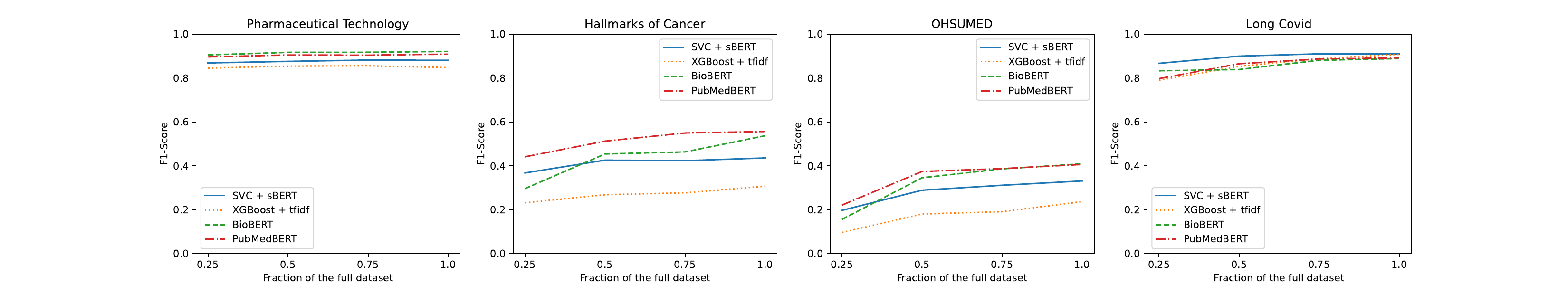}
    \caption{Task 2 (TC). Achieved F1 scores are shown for the test sets when training data is reduced.}
    \Description{Task 2 (TC). Achieved F1 scores are shown for the test sets when training data is reduced.}
    \label{fig:tc_reduced_training_data}
\end{figure*}

\begin{figure*}
    \centering
    \includegraphics[width=1.0\textwidth]{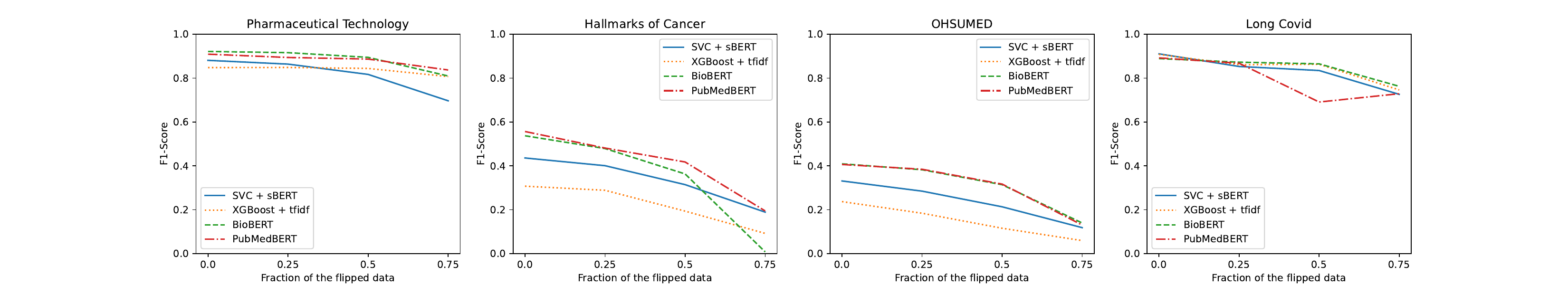}
    \caption{Task 2 (TC). Achieved F1 scores are shown for the test sets when flipping some training data.}
    \Description{Task 2 (TC). Achieved F1 scores are shown for the test sets when flipping some training data.}
    \label{fig:tc_flipping_training_data}
\end{figure*}

%\todo{MK: update numbers}
\begin{table}[t]
    \centering
    \resizebox{.45\textwidth}{!}{
    \begin{tabular}{l|rrrr}
        Model & Training  & HS & Application & ET PubMed \\
        \toprule

        \multicolumn{5}{c}{Traditional Classification Models} \\
        \midrule
        SVC + tfidf & 57 min 28 s & 42 h 41 min 7 s     & 9.79e-04 s & 9 h 56 min 40.26 s\\
        SVC + sBERT & 1 h 0 min 5 s & 53 h 17 min 10 s  & 9.70e-04 s & 9 h 50 min 47.59 s\\
        XGBoost + tfidf & \textbf{9 s} & 6 min 51 s     & \textbf{5.63e-05 s} & \textbf{34 min 17.15 s}\\
        XGBoost + sBERT & \textbf{9 s} & \textbf{6 min 37 s} & 7.00e-05 s & 42 min 39.25 s\\
        Random Forest + tfidf & 2 min 51 s & 7 min 38 s & 7.96e-05 s & 48 min 28.16 s\\
        Random Forest + sBERT & 3 min 27 s & 8 min 24 s & 7.68e-05 s & 46 min 48.54 s\\
       % \midrule
       % \multicolumn{5}{c}{Language Models on CPU} \\
       % \midrule
       % BERT & & & & \\
       % RoBERTa & & &  &\\
       % XLNet & & &  &\\
       % BioBERT & & &  &\\
       % BioLinkBERT & & & & \\
       % PubMedBERT & & & & \\
        \midrule
        \multicolumn{5}{c}{Language Models on GPU} \\
        \midrule
        BERT & 27 min 24 s & 12 h 14 min 4 s & 4.02e-03 s & 1 d 16 h 52 min 12.99 s \\
        RoBERTa & \textbf{9 min 48 s} & 8 h 40 min 25 s & \textbf{3.92e-03 s} & \textbf{1 d 15 h 49 min 11.95 s} \\
        XLNet & 12 min 12 s & 22 h 6 min 47 s & 1.15e-02 s & 4 d 20 h 34 min 6.06 s \\
        BioBERT & 10 min 6 s & 10 h 44 min 4 s & 3.97e-03 s & 1 d 16 h 18 min 5.54 s \\
        BioLinkBERT & 9 min 53 s & \textbf{5 h 57 min 51} s & 4.00e-03 s & 1 d 16 h 34 min 23.02 s \\
        PubMedBERT & 28 min 29 s & 9 h 30 min 4 s & 3.99e-03 s & 1 d 16 h 33 min 50.04 s \\
    \end{tabular}
    }
    \caption{Task 2 (TC). Runtimes were measured on the Pharmaceutical Technology benchmark. The training and HS time is reported in total, whereas the application time is normalized by the number of sentences in the test set. ET PubMed is the estimated time when applying the final model to the PubMed collection (37M documents).}
    \label{tab:tc_time}
\end{table}

%\todo{MK: insert largest benchmark in caption}

\paragraph{Application Costs}
Next, we measured the required time for training, hyperparameter search (HS), and application time normalized by the number of documents in the test set.
Again, we used the number of publications in PubMed (37M) to calculate the estimated time (ET PubMed) for a model application to the whole PubMed collection.
We measured the runtimes on the Pharmaceutical Technology benchmark as it was the largest text classification benchmark in our setup. 
The measured times are listed in Table~\ref{tab:tc_time}.
GPU application times are again, similarly to relation extraction, measured on an upscaled data set (1M abstracts) but still normalized.
In the Pharmaceutical Technology benchmark, XGBoost models were efficient, with quick training times and moderate F1 scores, processing PubMed relatively quickly. SVC models achieved slightly higher F1 scores but required significantly more time for PubMed processing. Using our GPU, LMs like BioBERT and BioLinkBERT offered a suitable mixture of quality and efficiency, with higher F1 scores and acceptable processing times. Again, XLNet was slower across all stages than BERT-based models.
Note that we skipped measuring LMs on CPU here because times would be very high and similar to the relation extraction task.

In summary, traditional models like XGBoost offered quick training, tuning and application times. However, LMs such as BioBERT and PubMedBERT, when using GPUs, offered still acceptable application times but achieved higher F1 scores. They still processed large datasets like PubMed very fast, making them better suited for tasks requiring high-quality classification.

\begin{table}
    \centering
    \resizebox{.45\textwidth}{!}{
    \begin{tabular}{l|rrr|rrr|rrr|rrr}
        Model &  \multicolumn{3}{c}{Hallmark}  & \multicolumn{3}{c}{Ohsumed}  & \multicolumn{3}{c}{Long COVID}  & \multicolumn{3}{c}{Pharm. Tech.} \\
       &  P & R & F1 & P & R & F1& P & R & F1& P & R & F1 \\
        \toprule
        \multicolumn{13}{c}{SingleTask Learning} \\
        \midrule
        SVC + tfidf & 0.36 & 0.67 & 0.44 & 0.31 & 0.39 & 0.33 & \textbf{0.92} & 0.90 & \textbf{0.91} & 0.87 & 0.89 & 0.88  \\
        XGBoost + tfidf & 0.27 & 0.57 & 0.31 & 0.25 & 0.29 & 0.24 & 0.88 & \textbf{0.94} & \textbf{0.91} & 0.86 & 0.84 & 0.85  \\
        PubMedBERT & \textbf{0.46} & \textbf{0.82} & \textbf{0.56} & \textbf{0.38} & \textbf{0.51} & \textbf{0.41} & 0.87 & 0.91 & 0.89 & 0.90 & 0.92 & 0.91  \\
        BioBERT & 0.44 & 0.8 & 0.54 & \textbf{0.38} & \textbf{0.51} & \textbf{0.41} & 0.90 & 0.88 & 0.89 & \textbf{0.91} & \textbf{0.93} & \textbf{0.92}  \\

        \midrule
        \multicolumn{13}{c}{MultiTask Learning} \\
        \midrule
        SVC + tfidf & 0.08 & 0.0 & 0.0 & 0.0 & 0.0 & 0.0 & 0.25 & 0.5 & 0.33 & 0.0 & 0.0 & 0.0  \\
        XGBoost + tfidf & 0.07 & 0.13 & 0.07 & 0.14 & 0.11 & 0.09 & 0.14 & 0.16 & 0.14 & 0.04 & 0.01 & 0.02  \\
        PubMedBERT & 0.31 & 0.63 & 0.38 & 0.31 & 0.42 & 0.33 & \textbf{0.76} & \textbf{0.72} & \textbf{0.71} & 0.04 & \textbf{0.03} & \textbf{0.03}  \\
        BioBERT & \textbf{0.34} & \textbf{0.68} & \textbf{0.41} & \textbf{0.36} & \textbf{0.49} & \textbf{0.39} & 0.44 & 0.36 & 0.29 & \textbf{0.05} & \textbf{0.03} & \textbf{0.03}  \\
       
    \end{tabular}
    }
    \caption{Task 2 (TC). We compared precision, recall and F1 of SingleTask vs. MultiTask learning on the benchmarks.} 
    \label{tab:tc_stvsmt}
\end{table}

\subsection{RQ2: System Architecture}
We again designed two system architectures: 1) SingleTask learning, i.e., a model for each document classification system. 2) MultiTask learning, i.e., combining the training data of every benchmark to train a single model.
Note that in practice, a document might have several classes which is not reflected in our evaluation setup (single class prediction). 
The comparison between both strategies is shown in Table~\ref{tab:tc_stvsmt}. Again, we compared the top two shallow models and the top two LMs based on their classification quality. The MultiTask Learning results showed issues across all four biomedical benchmarks. It resulted in much lower F1 scores, with models struggling to generalize across different datasets. Again, this is likely due to the small size of the training data after down-sampling. As predicted, LMs outperformed shallow models. Notably SingleTask strategy significantly outperformed MultiTask Learning, with models achieving better performance when trained individually.

\subsection{RQ3: Data Labeling}
Testing different strategies to label training data for the text classification task remains challenging: Distantly supervised approaches do obviously not work (if a knowledge base would store classes for documents, this would itself be a benchmark already).
It was also challenging to use LLMs to re-label training, similar to what we did for the relation extraction task. 
How should we create sophisticated prompts that tell an LLM whether an abstract should be about a certain topic?
Suitable examples must be found for each class when restricting prompts to examples (as in few-show prompting).
When using example terms for each class in prompts, the LLM will very likely decide whether some of the example terms are present.
A classification approach based on regular expression would then be sufficient for this scenario.
In brief, distant supervision and LLM-driven labeling cannot be applied to the text classification task without exhausting engineering (which was out of our scope).  
However, we decided not to skip the experiment completely. 
Instead, we tested two strategies to introduce noise in the training data set.
We assume that this noise could reflect some bad labeling strategy.
This assumption has problems as our noise will be introduced randomly, not by some bias/systematic error in the labeling.

Our first strategy reduces the number of available training data by random, i.e., we randomly sampled the training data down to specific fractions of the original dataset, such as 100\%, 75\%, 50\%, and 25\%. 
We then trained our best-performing model configurations on that reduced training data set and evaluated their F1 scores on the original test data set.
Figure~\ref{fig:tc_reduced_training_data} shows the impact of reducing training data on model performance across different benchmarks. In the Pharmaceutical Technology benchmark, all models showed stable performance despite data reduction. However, in the Hallmarks of Cancer and OHSUMED benchmarks, F1 scores increased significantly with more training data, which shows a stronger dependence on a larger dataset for improved performance. LMs benefit from increased training data size, consistently outperforming traditional models in these four benchmarks. Overall, LMs can manage data scarcity better than traditional models.

The second strategy introduces noise into the training data by randomly flipping a certain degree of labels. The flipping is done by randomly sampling the balanced train set by a defined flip value (e.g., 25 \%) and shuffling the labels within this subset. The flipped subset is then reintegrated with the original dataset, maintaining the overall data structure while introducing controlled label variability.

Again, we trained models on those data sets and measured their performance based on the original test sets.
Figure~\ref{fig:tc_flipping_training_data} visualize the results. In all the benchmarks, all models experienced a decline in F1 scores as the level of label noise increases. In the OHSUMED benchmark, the LMs showed greater resilience to this noise, maintaining higher performance than traditional models. However, in the Hallmarks of Cancer benchmark, BioBERT performed worst among all models when 75\% of the data is flipped, indicating a higher sensitivity to label noise in this specific scenario. This shows that LMs can still be vulnerable under extreme noise conditions.

\section{Conclusion}
While some of our findings were expected, e.g., that LMs are more robust and accurate on classification tasks, our paper contributes a library perspective when applying them. 
First, we compared different models in terms of their training and application times. 
GPUs are a must-have when working with LMs. Shallow models like SVC/XGBoost may still worth using. 
Second, LLMs can label training data with a moderate quality and costs. However, the overall classification quality is then decreased.  
In conclusion, we examined how to tackle supervised text processing within a digital library. We shared our code so other libraries could build upon our findings. Further investigation could shed more light on generating training data or finding more reliable MultiTask learning setups.

\section*{Acknowledgments}
Supported by the Deutsche Forschungsgemeinschaft (DFG, German Research Foundation): PubPharm – the Specialized Information Service for Pharmacy (Gepris 267140244).

\bibliographystyle{ACM-Reference-Format}
\bibliography{references}


\begin{thebibliography}{47}

%%% the Web address.

\ifx \showCODEN    \undefined \def \showCODEN     #1{\unskip}     \fi
\ifx \showDOI      \undefined \def \showDOI       #1{#1}\fi
\ifx \showISBNx    \undefined \def \showISBNx     #1{\unskip}     \fi
\ifx \showISBNxiii \undefined \def \showISBNxiii  #1{\unskip}     \fi
\ifx \showISSN     \undefined \def \showISSN      #1{\unskip}     \fi
\ifx \showLCCN     \undefined \def \showLCCN      #1{\unskip}     \fi
\ifx \shownote     \undefined \def \shownote      #1{#1}          \fi
\ifx \showarticletitle \undefined \def \showarticletitle #1{#1}   \fi
\ifx \showURL      \undefined \def \showURL       {\relax}        \fi
% invisible to TeX
\providecommand\bibfield[2]{#2}
\providecommand\bibinfo[2]{#2}
\providecommand\natexlab[1]{#1}
\providecommand\showeprint[2][]{arXiv:#2}

\bibitem[AI(2024)]%
        {gpt4o}
\bibfield{author}{\bibinfo{person}{Open AI}.} \bibinfo{year}{2024}\natexlab{}.
\newblock \bibinfo{title}{Hello GPT-4o}.
\newblock \bibinfo{howpublished}{\url{https://openai.com/index/hello-gpt-4o/}}.
\newblock
\newblock
\shownote{Accessed: 20 July 2024}.


\bibitem[AI@Meta(2024)]%
        {llama3}
\bibfield{author}{\bibinfo{person}{AI@Meta}.} \bibinfo{year}{2024}\natexlab{}.
\newblock \bibinfo{title}{Llama 3 Model Card}.
\newblock \bibinfo{howpublished}{\url{https://github.com/meta-llama/llama3/blob/main/MODEL_CARD.md}}.
\newblock
\newblock
\shownote{Accessed: 20 July 2024}.


\bibitem[Baker et~al\mbox{.}(2017)]%
        {DBLP:journals/bioinformatics/BakerASPGHSK17}
\bibfield{author}{\bibinfo{person}{Simon Baker}, \bibinfo{person}{Imran Ali}, \bibinfo{person}{Ilona Silins}, \bibinfo{person}{Sampo Pyysalo}, \bibinfo{person}{Yufan Guo}, \bibinfo{person}{Johan H{\"{o}}gberg}, \bibinfo{person}{Ulla Stenius}, {and} \bibinfo{person}{Anna Korhonen}.} \bibinfo{year}{2017}\natexlab{}.
\newblock \showarticletitle{Cancer Hallmarks Analytics Tool {(CHAT):} a text mining approach to organize and evaluate scientific literature on cancer}.
\newblock \bibinfo{journal}{\emph{Bioinform.}} \bibinfo{volume}{33}, \bibinfo{number}{24} (\bibinfo{year}{2017}), \bibinfo{pages}{3973--3981}.
\newblock
\urldef\tempurl%
\url{https://doi.org/10.1093/BIOINFORMATICS/BTX454}
\showDOI{\tempurl}


\bibitem[Baker et~al\mbox{.}(2016a)]%
        {DBLP:conf/coling/BakerKP16}
\bibfield{author}{\bibinfo{person}{Simon Baker}, \bibinfo{person}{Anna Korhonen}, {and} \bibinfo{person}{Sampo Pyysalo}.} \bibinfo{year}{2016}\natexlab{a}.
\newblock \showarticletitle{Cancer Hallmark Text Classification Using Convolutional Neural Networks}. In \bibinfo{booktitle}{\emph{Proceedings of the Fifth Workshop on Building and Evaluating Resources for Biomedical Text Mining, BioTxtM@COLING 2016, Osaka, Japan, December 12, 2016}}. \bibinfo{publisher}{The {COLING} 2016 Organizing Committee}, \bibinfo{pages}{1--9}.
\newblock
\urldef\tempurl%
\url{https://aclanthology.org/W16-5101/}
\showURL{%
\tempurl}


\bibitem[Baker et~al\mbox{.}(2016b)]%
        {DBLP:journals/bioinformatics/BakerSGAHSK16}
\bibfield{author}{\bibinfo{person}{Simon Baker}, \bibinfo{person}{Ilona Silins}, \bibinfo{person}{Yufan Guo}, \bibinfo{person}{Imran Ali}, \bibinfo{person}{Johan H{\"{o}}gberg}, \bibinfo{person}{Ulla Stenius}, {and} \bibinfo{person}{Anna Korhonen}.} \bibinfo{year}{2016}\natexlab{b}.
\newblock \showarticletitle{Automatic semantic classification of scientific literature according to the hallmarks of cancer}.
\newblock \bibinfo{journal}{\emph{Bioinform.}} \bibinfo{volume}{32}, \bibinfo{number}{3} (\bibinfo{year}{2016}), \bibinfo{pages}{432--440}.
\newblock
\urldef\tempurl%
\url{https://doi.org/10.1093/BIOINFORMATICS/BTV585}
\showDOI{\tempurl}


\bibitem[Chia et~al\mbox{.}(2022)]%
        {DBLP:conf/acl/ChiaBPS22}
\bibfield{author}{\bibinfo{person}{Yew~Ken Chia}, \bibinfo{person}{Lidong Bing}, \bibinfo{person}{Soujanya Poria}, {and} \bibinfo{person}{Luo Si}.} \bibinfo{year}{2022}\natexlab{}.
\newblock \showarticletitle{RelationPrompt: Leveraging Prompts to Generate Synthetic Data for Zero-Shot Relation Triplet Extraction}. In \bibinfo{booktitle}{\emph{Findings of the Association for Computational Linguistics: {ACL} 2022, Dublin, Ireland, May 22-27, 2022}}. \bibinfo{publisher}{Association for Computational Linguistics}, \bibinfo{pages}{45--57}.
\newblock
\urldef\tempurl%
\url{https://doi.org/10.18653/V1/2022.FINDINGS-ACL.5}
\showDOI{\tempurl}


\bibitem[Davis et~al\mbox{.}(2023)]%
        {CTD:DBLP:journals/nar/DavisWJSWM23}
\bibfield{author}{\bibinfo{person}{Allan~Peter Davis}, \bibinfo{person}{Thomas~C. Wiegers}, \bibinfo{person}{Robin~J. Johnson}, \bibinfo{person}{Daniela Sciaky}, \bibinfo{person}{Jolene Wiegers}, {and} \bibinfo{person}{Carolyn~J. Mattingly}.} \bibinfo{year}{2023}\natexlab{}.
\newblock \showarticletitle{Comparative Toxicogenomics Database {(CTD):} update 2023}.
\newblock \bibinfo{journal}{\emph{Nucleic Acids Res.}} \bibinfo{volume}{51}, \bibinfo{number}{{D1}} (\bibinfo{year}{2023}), \bibinfo{pages}{1257--1262}.
\newblock
\urldef\tempurl%
\url{https://doi.org/10.1093/NAR/GKAC833}
\showDOI{\tempurl}


\bibitem[Detroja et~al\mbox{.}(2023)]%
        {DBLP:journals/iswa/DetrojaBB23}
\bibfield{author}{\bibinfo{person}{Kartik Detroja}, \bibinfo{person}{C.~K. Bhensdadia}, {and} \bibinfo{person}{Brijesh~S. Bhatt}.} \bibinfo{year}{2023}\natexlab{}.
\newblock \showarticletitle{A survey on Relation Extraction}.
\newblock \bibinfo{journal}{\emph{Intell. Syst. Appl.}}  \bibinfo{volume}{19} (\bibinfo{year}{2023}), \bibinfo{pages}{200244}.
\newblock
\urldef\tempurl%
\url{https://doi.org/10.1016/J.ISWA.2023.200244}
\showDOI{\tempurl}


\bibitem[Devlin et~al\mbox{.}(2019)]%
        {devlin2019bert}
\bibfield{author}{\bibinfo{person}{Jacob Devlin}, \bibinfo{person}{Ming-Wei Chang}, \bibinfo{person}{Kenton Lee}, {and} \bibinfo{person}{Kristina Toutanova}.} \bibinfo{year}{2019}\natexlab{}.
\newblock \showarticletitle{{BERT}: Pre-training of Deep Bidirectional Transformers for Language Understanding}. In \bibinfo{booktitle}{\emph{Proceedings of the 2019 Conference of the North {A}merican Chapter of the Association for Computational Linguistics: Human Language Technologies, Volume 1 (Long and Short Papers)}}. \bibinfo{publisher}{Association for Computational Linguistics}, \bibinfo{address}{Minneapolis, Minnesota}, \bibinfo{pages}{4171--4186}.
\newblock
\urldef\tempurl%
\url{https://doi.org/10.18653/v1/N19-1423}
\showDOI{\tempurl}


\bibitem[D{\"{u}}ring et~al\mbox{.}(2021)]%
        {impresso}
\bibfield{author}{\bibinfo{person}{Marten D{\"{u}}ring}, \bibinfo{person}{Roman Kalyakin}, \bibinfo{person}{Estelle Bunout}, {and} \bibinfo{person}{Daniele Guido}.} \bibinfo{year}{2021}\natexlab{}.
\newblock \showarticletitle{Impresso Inspect and Compare. Visual Comparison of Semantically Enriched Historical Newspaper Articles}.
\newblock \bibinfo{journal}{\emph{Inf.}} \bibinfo{volume}{12}, \bibinfo{number}{9} (\bibinfo{year}{2021}), \bibinfo{pages}{348}.
\newblock
\urldef\tempurl%
\url{https://doi.org/10.3390/info12090348}
\showDOI{\tempurl}


\bibitem[Groeneveld et~al\mbox{.}(2024)]%
        {olmo}
\bibfield{author}{\bibinfo{person}{Dirk Groeneveld}, \bibinfo{person}{Iz Beltagy}, \bibinfo{person}{Evan~Pete Walsh}, \bibinfo{person}{Akshita Bhagia}, \bibinfo{person}{Rodney Kinney}, {and} \bibinfo{person}{Oyvind~Tafjord et al.}} \bibinfo{year}{2024}\natexlab{}.
\newblock \showarticletitle{OLMo: Accelerating the Science of Language Models}. In \bibinfo{booktitle}{\emph{Proceedings of the 62nd Annual Meeting of the Association for Computational Linguistics (Volume 1: Long Papers), {ACL} 2024, Bangkok, Thailand, August 11-16, 2024}}. \bibinfo{publisher}{Association for Computational Linguistics}, \bibinfo{pages}{15789--15809}.
\newblock
\urldef\tempurl%
\url{https://doi.org/10.18653/V1/2024.ACL-LONG.841}
\showDOI{\tempurl}


\bibitem[Gu et~al\mbox{.}(2022)]%
        {DBLP:journals/health/GuTCLULNGP22}
\bibfield{author}{\bibinfo{person}{Yu Gu}, \bibinfo{person}{Robert Tinn}, \bibinfo{person}{Hao Cheng}, \bibinfo{person}{Michael Lucas}, \bibinfo{person}{Naoto Usuyama}, \bibinfo{person}{Xiaodong Liu}, \bibinfo{person}{Tristan Naumann}, \bibinfo{person}{Jianfeng Gao}, {and} \bibinfo{person}{Hoifung Poon}.} \bibinfo{year}{2022}\natexlab{}.
\newblock \showarticletitle{Domain-Specific Language Model Pretraining for Biomedical Natural Language Processing}.
\newblock \bibinfo{journal}{\emph{{ACM} Trans. Comput. Heal.}} \bibinfo{volume}{3}, \bibinfo{number}{1} (\bibinfo{year}{2022}), \bibinfo{pages}{2:1--2:23}.
\newblock
\urldef\tempurl%
\url{https://doi.org/10.1145/3458754}
\showDOI{\tempurl}


\bibitem[Herrero{-}Zazo et~al\mbox{.}(2013)]%
        {DDI:DBLP:journals/jbi/Herrero-ZazoSMD13}
\bibfield{author}{\bibinfo{person}{Mar{\'{\i}}a Herrero{-}Zazo}, \bibinfo{person}{Isabel Segura{-}Bedmar}, \bibinfo{person}{Paloma Mart{\'{\i}}nez}, {and} \bibinfo{person}{Thierry Declerck}.} \bibinfo{year}{2013}\natexlab{}.
\newblock \showarticletitle{The {DDI} corpus: An annotated corpus with pharmacological substances and drug-drug interactions}.
\newblock \bibinfo{journal}{\emph{J. Biomed. Informatics}} \bibinfo{volume}{46}, \bibinfo{number}{5} (\bibinfo{year}{2013}), \bibinfo{pages}{914--920}.
\newblock
\urldef\tempurl%
\url{https://doi.org/10.1016/J.JBI.2013.07.011}
\showDOI{\tempurl}


\bibitem[Hersh et~al\mbox{.}(1994)]%
        {DBLP:conf/sigir/HershBLH94}
\bibfield{author}{\bibinfo{person}{William~R. Hersh}, \bibinfo{person}{Chris Buckley}, \bibinfo{person}{T.~J. Leone}, {and} \bibinfo{person}{David~H. Hickam}.} \bibinfo{year}{1994}\natexlab{}.
\newblock \showarticletitle{{OHSUMED:} An Interactive Retrieval Evaluation and New Large Test Collection for Research}. In \bibinfo{booktitle}{\emph{Proceedings of the 17th Annual International {ACM-SIGIR} Conference on Research and Development in Information Retrieval. Dublin, Ireland, 3-6 July 1994 (Special Issue of the {SIGIR} Forum)}}. \bibinfo{publisher}{ACM/Springer}, \bibinfo{pages}{192--201}.
\newblock
\urldef\tempurl%
\url{https://doi.org/10.1007/978-1-4471-2099-5\_20}
\showDOI{\tempurl}


\bibitem[Joachims(1998)]%
        {DBLP:conf/ecml/Joachims98}
\bibfield{author}{\bibinfo{person}{Thorsten Joachims}.} \bibinfo{year}{1998}\natexlab{}.
\newblock \showarticletitle{Text Categorization with Support Vector Machines: Learning with Many Relevant Features}. In \bibinfo{booktitle}{\emph{Machine Learning: ECML-98, 10th European Conference on Machine Learning, Chemnitz, Germany, April 21-23, 1998, Proceedings}} \emph{(\bibinfo{series}{Lecture Notes in Computer Science}, Vol.~\bibinfo{volume}{1398})}. \bibinfo{publisher}{Springer}, \bibinfo{pages}{137--142}.
\newblock
\urldef\tempurl%
\url{https://doi.org/10.1007/BFB0026683}
\showDOI{\tempurl}


\bibitem[Josifoski et~al\mbox{.}(2023)]%
        {DBLP:conf/emnlp/JosifoskiSP023}
\bibfield{author}{\bibinfo{person}{Martin Josifoski}, \bibinfo{person}{Marija Sakota}, \bibinfo{person}{Maxime Peyrard}, {and} \bibinfo{person}{Robert West}.} \bibinfo{year}{2023}\natexlab{}.
\newblock \showarticletitle{Exploiting Asymmetry for Synthetic Training Data Generation: SynthIE and the Case of Information Extraction}. In \bibinfo{booktitle}{\emph{Proceedings of the 2023 Conference on Empirical Methods in Natural Language Processing, {EMNLP} 2023, Singapore, December 6-10, 2023}}. \bibinfo{publisher}{Association for Computational Linguistics}, \bibinfo{pages}{1555--1574}.
\newblock
\urldef\tempurl%
\url{https://doi.org/10.18653/V1/2023.EMNLP-MAIN.96}
\showDOI{\tempurl}


\bibitem[Kabongo et~al\mbox{.}(2024)]%
        {DBLP:journals/jodl/KabongoDA24}
\bibfield{author}{\bibinfo{person}{Salomon Kabongo}, \bibinfo{person}{Jennifer D'Souza}, {and} \bibinfo{person}{S{\"{o}}ren Auer}.} \bibinfo{year}{2024}\natexlab{}.
\newblock \showarticletitle{ORKG-Leaderboards: a systematic workflow for mining leaderboards as a knowledge graph}.
\newblock \bibinfo{journal}{\emph{Int. J. Digit. Libr.}} \bibinfo{volume}{25}, \bibinfo{number}{1} (\bibinfo{year}{2024}), \bibinfo{pages}{41--54}.
\newblock
\urldef\tempurl%
\url{https://doi.org/10.1007/S00799-023-00366-1}
\showDOI{\tempurl}


\bibitem[Kilicoglu et~al\mbox{.}(2012)]%
        {DBLP:journals/bioinformatics/KilicogluSFRR12}
\bibfield{author}{\bibinfo{person}{Halil Kilicoglu}, \bibinfo{person}{Dongwook Shin}, \bibinfo{person}{Marcelo Fiszman}, \bibinfo{person}{Graciela Rosemblat}, {and} \bibinfo{person}{Thomas~C. Rindflesch}.} \bibinfo{year}{2012}\natexlab{}.
\newblock \showarticletitle{SemMedDB: a PubMed-scale repository of biomedical semantic predications}.
\newblock \bibinfo{journal}{\emph{Bioinform.}} \bibinfo{volume}{28}, \bibinfo{number}{23} (\bibinfo{year}{2012}), \bibinfo{pages}{3158--3160}.
\newblock
\urldef\tempurl%
\url{https://doi.org/10.1093/BIOINFORMATICS/BTS591}
\showDOI{\tempurl}


\bibitem[Krallinger et~al\mbox{.}(2017)]%
        {ChemProt:krallinger2017overview}
\bibfield{author}{\bibinfo{person}{Martin Krallinger}, \bibinfo{person}{Obdulia Rabal}, \bibinfo{person}{Saber~A Akhondi}, \bibinfo{person}{Mart{\i}n~P{\'e}rez P{\'e}rez}, \bibinfo{person}{Jes{\'u}s Santamar{\'\i}a}, \bibinfo{person}{Gael~P{\'e}rez Rodr{\'\i}guez}, \bibinfo{person}{Georgios Tsatsaronis}, \bibinfo{person}{Ander Intxaurrondo}, \bibinfo{person}{Jos{\'e}~Antonio L{\'o}pez}, \bibinfo{person}{Umesh Nandal}, {et~al\mbox{.}}} \bibinfo{year}{2017}\natexlab{}.
\newblock \showarticletitle{Overview of the BioCreative VI chemical-protein interaction Track}. In \bibinfo{booktitle}{\emph{Proceedings of the sixth BioCreative challenge evaluation workshop}}, Vol.~\bibinfo{volume}{1}. \bibinfo{pages}{141--146}.
\newblock


\bibitem[Kroll et~al\mbox{.}(2024a)]%
        {DBLP:journals/jodl/KrollPKKRB24}
\bibfield{author}{\bibinfo{person}{Hermann Kroll}, \bibinfo{person}{Jan Pirklbauer}, \bibinfo{person}{Jan{-}Christoph Kalo}, \bibinfo{person}{Morris Kunz}, \bibinfo{person}{Johannes Ruthmann}, {and} \bibinfo{person}{Wolf{-}Tilo Balke}.} \bibinfo{year}{2024}\natexlab{a}.
\newblock \showarticletitle{A discovery system for narrative query graphs: entity-interaction-aware document retrieval}.
\newblock \bibinfo{journal}{\emph{Int. J. Digit. Libr.}} \bibinfo{volume}{25}, \bibinfo{number}{1} (\bibinfo{year}{2024}), \bibinfo{pages}{3--24}.
\newblock
\urldef\tempurl%
\url{https://doi.org/10.1007/S00799-023-00356-3}
\showDOI{\tempurl}


\bibitem[Kroll et~al\mbox{.}(2022)]%
        {DBLP:conf/jcdl/KrollPPB22a}
\bibfield{author}{\bibinfo{person}{Hermann Kroll}, \bibinfo{person}{Jan Pirklbauer}, \bibinfo{person}{Florian Pl{\"{o}}tzky}, {and} \bibinfo{person}{Wolf{-}Tilo Balke}.} \bibinfo{year}{2022}\natexlab{}.
\newblock \showarticletitle{A library perspective on nearly-unsupervised information extraction workflows in digital libraries}. In \bibinfo{booktitle}{\emph{{JCDL} '22: The {ACM/IEEE} Joint Conference on Digital Libraries in 2022, Cologne, Germany, June 20 - 24, 2022}}. \bibinfo{publisher}{{ACM}}, \bibinfo{pages}{35}.
\newblock
\urldef\tempurl%
\url{https://doi.org/10.1145/3529372.3530924}
\showDOI{\tempurl}


\bibitem[Kroll et~al\mbox{.}(2024b)]%
        {DBLP:journals/jodl/KrollPPB24}
\bibfield{author}{\bibinfo{person}{Hermann Kroll}, \bibinfo{person}{Jan Pirklbauer}, \bibinfo{person}{Florian Pl{\"{o}}tzky}, {and} \bibinfo{person}{Wolf{-}Tilo Balke}.} \bibinfo{year}{2024}\natexlab{b}.
\newblock \showarticletitle{A detailed library perspective on nearly unsupervised information extraction workflows in digital libraries}.
\newblock \bibinfo{journal}{\emph{Int. J. Digit. Libr.}} \bibinfo{volume}{25}, \bibinfo{number}{2} (\bibinfo{year}{2024}), \bibinfo{pages}{401--425}.
\newblock
\urldef\tempurl%
\url{https://doi.org/10.1007/S00799-023-00368-Z}
\showDOI{\tempurl}


\bibitem[Kumar and Sharaff(2024)]%
        {DBLP:journals/cj/KumarS24}
\bibfield{author}{\bibinfo{person}{Ashutosh Kumar} {and} \bibinfo{person}{Aakanksha Sharaff}.} \bibinfo{year}{2024}\natexlab{}.
\newblock \showarticletitle{SnorkelPlus: {A} Novel Approach for Identifying Relationships Among Biomedical Entities Within Abstracts}.
\newblock \bibinfo{journal}{\emph{Comput. J.}} \bibinfo{volume}{67}, \bibinfo{number}{3} (\bibinfo{year}{2024}), \bibinfo{pages}{1187--1200}.
\newblock
\urldef\tempurl%
\url{https://doi.org/10.1093/COMJNL/BXAD051}
\showDOI{\tempurl}


\bibitem[Labrak et~al\mbox{.}(2024)]%
        {DBLP:conf/acl/LabrakBMGRD24}
\bibfield{author}{\bibinfo{person}{Yanis Labrak}, \bibinfo{person}{Adrien Bazoge}, \bibinfo{person}{Emmanuel Morin}, \bibinfo{person}{Pierre{-}Antoine Gourraud}, \bibinfo{person}{Mickael Rouvier}, {and} \bibinfo{person}{Richard Dufour}.} \bibinfo{year}{2024}\natexlab{}.
\newblock \showarticletitle{BioMistral: {A} Collection of Open-Source Pretrained Large Language Models for Medical Domains}. In \bibinfo{booktitle}{\emph{Findings of the Association for Computational Linguistics, {ACL} 2024, Bangkok, Thailand and virtual meeting, August 11-16, 2024}}. \bibinfo{publisher}{Association for Computational Linguistics}, \bibinfo{pages}{5848--5864}.
\newblock
\urldef\tempurl%
\url{https://doi.org/10.18653/V1/2024.FINDINGS-ACL.348}
\showDOI{\tempurl}


\bibitem[Lai et~al\mbox{.}(2023)]%
        {DBLP:journals/jbi/LaiWLCL23}
\bibfield{author}{\bibinfo{person}{Po{-}Ting Lai}, \bibinfo{person}{Chih{-}Hsuan Wei}, \bibinfo{person}{Ling Luo}, \bibinfo{person}{Qingyu Chen}, {and} \bibinfo{person}{Zhiyong Lu}.} \bibinfo{year}{2023}\natexlab{}.
\newblock \showarticletitle{BioREx: Improving biomedical relation extraction by leveraging heterogeneous datasets}.
\newblock \bibinfo{journal}{\emph{J. Biomed. Informatics}}  \bibinfo{volume}{146} (\bibinfo{year}{2023}), \bibinfo{pages}{104487}.
\newblock
\urldef\tempurl%
\url{https://doi.org/10.1016/J.JBI.2023.104487}
\showDOI{\tempurl}


\bibitem[Langnickel et~al\mbox{.}(2022)]%
        {DBLP:journals/biodb/LangnickelDHDF22}
\bibfield{author}{\bibinfo{person}{Lisa Langnickel}, \bibinfo{person}{Johannes Darms}, \bibinfo{person}{Katharina Heldt}, \bibinfo{person}{Denise Ducks}, {and} \bibinfo{person}{Juliane Fluck}.} \bibinfo{year}{2022}\natexlab{}.
\newblock \showarticletitle{Continuous development of the semantic search engine \emph{preVIEW}: from {COVID-19} to long {COVID}}.
\newblock \bibinfo{journal}{\emph{Database J. Biol. Databases Curation}} \bibinfo{volume}{2022}, \bibinfo{number}{2022} (\bibinfo{year}{2022}).
\newblock
\urldef\tempurl%
\url{https://doi.org/10.1093/DATABASE/BAAC048}
\showDOI{\tempurl}


\bibitem[Leaman and Lu(2016)]%
        {DBLP:journals/bioinformatics/LeamanL16}
\bibfield{author}{\bibinfo{person}{Robert Leaman} {and} \bibinfo{person}{Zhiyong Lu}.} \bibinfo{year}{2016}\natexlab{}.
\newblock \showarticletitle{TaggerOne: joint named entity recognition and normalization with semi-Markov Models}.
\newblock \bibinfo{journal}{\emph{Bioinform.}} \bibinfo{volume}{32}, \bibinfo{number}{18} (\bibinfo{year}{2016}), \bibinfo{pages}{2839--2846}.
\newblock
\urldef\tempurl%
\url{https://doi.org/10.1093/BIOINFORMATICS/BTW343}
\showDOI{\tempurl}


\bibitem[Lee et~al\mbox{.}(2020)]%
        {lee2020biobert}
\bibfield{author}{\bibinfo{person}{Jinhyuk Lee}, \bibinfo{person}{Wonjin Yoon}, \bibinfo{person}{Sungdong Kim}, \bibinfo{person}{Donghyeon Kim}, \bibinfo{person}{Sunkyu Kim}, \bibinfo{person}{Chan~Ho So}, {and} \bibinfo{person}{Jaewoo Kang}.} \bibinfo{year}{2020}\natexlab{}.
\newblock \showarticletitle{BioBERT: a pre-trained biomedical language representation model for biomedical text mining}.
\newblock \bibinfo{journal}{\emph{Bioinform.}} \bibinfo{volume}{36}, \bibinfo{number}{4} (\bibinfo{year}{2020}), \bibinfo{pages}{1234--1240}.
\newblock
\urldef\tempurl%
\url{https://doi.org/10.1093/bioinformatics/btz682}
\showDOI{\tempurl}


\bibitem[Li et~al\mbox{.}(2022)]%
        {DBLP:journals/tist/LiPLXYSYH22}
\bibfield{author}{\bibinfo{person}{Qian Li}, \bibinfo{person}{Hao Peng}, \bibinfo{person}{Jianxin Li}, \bibinfo{person}{Congying Xia}, \bibinfo{person}{Renyu Yang}, \bibinfo{person}{Lichao Sun}, \bibinfo{person}{Philip~S. Yu}, {and} \bibinfo{person}{Lifang He}.} \bibinfo{year}{2022}\natexlab{}.
\newblock \showarticletitle{A Survey on Text Classification: From Traditional to Deep Learning}.
\newblock \bibinfo{journal}{\emph{{ACM} Trans. Intell. Syst. Technol.}} \bibinfo{volume}{13}, \bibinfo{number}{2} (\bibinfo{year}{2022}), \bibinfo{pages}{31:1--31:41}.
\newblock
\urldef\tempurl%
\url{https://doi.org/10.1145/3495162}
\showDOI{\tempurl}


\bibitem[Li et~al\mbox{.}(2023)]%
        {DBLP:conf/emnlp/LiZL023}
\bibfield{author}{\bibinfo{person}{Zhuoyan Li}, \bibinfo{person}{Hangxiao Zhu}, \bibinfo{person}{Zhuoran Lu}, {and} \bibinfo{person}{Ming Yin}.} \bibinfo{year}{2023}\natexlab{}.
\newblock \showarticletitle{Synthetic Data Generation with Large Language Models for Text Classification: Potential and Limitations}. In \bibinfo{booktitle}{\emph{Proceedings of the 2023 Conference on Empirical Methods in Natural Language Processing, {EMNLP} 2023, Singapore, December 6-10, 2023}}. \bibinfo{publisher}{Association for Computational Linguistics}, \bibinfo{pages}{10443--10461}.
\newblock
\urldef\tempurl%
\url{https://doi.org/10.18653/V1/2023.EMNLP-MAIN.647}
\showDOI{\tempurl}


\bibitem[Liu et~al\mbox{.}(2019)]%
        {liu2019roberta}
\bibfield{author}{\bibinfo{person}{Yinhan Liu}, \bibinfo{person}{Myle Ott}, \bibinfo{person}{Naman Goyal}, \bibinfo{person}{Jingfei Du}, \bibinfo{person}{Mandar Joshi}, \bibinfo{person}{Danqi Chen}, \bibinfo{person}{Omer Levy}, \bibinfo{person}{Mike Lewis}, \bibinfo{person}{Luke Zettlemoyer}, {and} \bibinfo{person}{Veselin Stoyanov}.} \bibinfo{year}{2019}\natexlab{}.
\newblock \showarticletitle{RoBERTa: {A} Robustly Optimized {BERT} Pretraining Approach}.
\newblock \bibinfo{journal}{\emph{CoRR}}  \bibinfo{volume}{abs/1907.11692} (\bibinfo{year}{2019}).
\newblock
\showeprint[arXiv]{1907.11692}
\urldef\tempurl%
\url{http://arxiv.org/abs/1907.11692}
\showURL{%
\tempurl}


\bibitem[Milosevic and Thielemann(2023)]%
        {DBLP:journals/ws/MilosevicT23}
\bibfield{author}{\bibinfo{person}{Nikola Milosevic} {and} \bibinfo{person}{Wolfgang Thielemann}.} \bibinfo{year}{2023}\natexlab{}.
\newblock \showarticletitle{Comparison of biomedical relationship extraction methods and models for knowledge graph creation}.
\newblock \bibinfo{journal}{\emph{J. Web Semant.}}  \bibinfo{volume}{75} (\bibinfo{year}{2023}), \bibinfo{pages}{100756}.
\newblock
\urldef\tempurl%
\url{https://doi.org/10.1016/J.WEBSEM.2022.100756}
\showDOI{\tempurl}


\bibitem[Peng et~al\mbox{.}(2024)]%
        {DBLP:journals/jbi/PengYSYCBW24}
\bibfield{author}{\bibinfo{person}{Cheng Peng}, \bibinfo{person}{Xi Yang}, \bibinfo{person}{Kaleb~E. Smith}, \bibinfo{person}{Zehao Yu}, \bibinfo{person}{Aokun Chen}, \bibinfo{person}{Jiang Bian}, {and} \bibinfo{person}{Yonghui Wu}.} \bibinfo{year}{2024}\natexlab{}.
\newblock \showarticletitle{Model tuning or prompt Tuning? a study of large language models for clinical concept and relation extraction}.
\newblock \bibinfo{journal}{\emph{J. Biomed. Informatics}}  \bibinfo{volume}{153} (\bibinfo{year}{2024}), \bibinfo{pages}{104630}.
\newblock
\urldef\tempurl%
\url{https://doi.org/10.1016/J.JBI.2024.104630}
\showDOI{\tempurl}


\bibitem[Ratner et~al\mbox{.}(2020)]%
        {DBLP:journals/vldb/RatnerBEFWR20}
\bibfield{author}{\bibinfo{person}{Alexander Ratner}, \bibinfo{person}{Stephen~H. Bach}, \bibinfo{person}{Henry~R. Ehrenberg}, \bibinfo{person}{Jason~A. Fries}, \bibinfo{person}{Sen Wu}, {and} \bibinfo{person}{Christopher R{\'{e}}}.} \bibinfo{year}{2020}\natexlab{}.
\newblock \showarticletitle{Snorkel: rapid training data creation with weak supervision}.
\newblock \bibinfo{journal}{\emph{{VLDB} J.}} \bibinfo{volume}{29}, \bibinfo{number}{2-3} (\bibinfo{year}{2020}), \bibinfo{pages}{709--730}.
\newblock
\urldef\tempurl%
\url{https://doi.org/10.1007/S00778-019-00552-1}
\showDOI{\tempurl}


\bibitem[Reimers and Gurevych(2019)]%
        {DBLP:conf/emnlp/ReimersG19}
\bibfield{author}{\bibinfo{person}{Nils Reimers} {and} \bibinfo{person}{Iryna Gurevych}.} \bibinfo{year}{2019}\natexlab{}.
\newblock \showarticletitle{Sentence-BERT: Sentence Embeddings using Siamese BERT-Networks}. In \bibinfo{booktitle}{\emph{Proceedings of the 2019 Conference on Empirical Methods in Natural Language Processing and the 9th International Joint Conference on Natural Language Processing, {EMNLP-IJCNLP} 2019, Hong Kong, China, November 3-7, 2019}}. \bibinfo{publisher}{Association for Computational Linguistics}, \bibinfo{pages}{3980--3990}.
\newblock
\urldef\tempurl%
\url{https://doi.org/10.18653/V1/D19-1410}
\showDOI{\tempurl}


\bibitem[Smirnova and Cudr{\'{e}}{-}Mauroux(2019)]%
        {DBLP:journals/csur/SmirnovaC19}
\bibfield{author}{\bibinfo{person}{Alisa Smirnova} {and} \bibinfo{person}{Philippe Cudr{\'{e}}{-}Mauroux}.} \bibinfo{year}{2019}\natexlab{}.
\newblock \showarticletitle{Relation Extraction Using Distant Supervision: {A} Survey}.
\newblock \bibinfo{journal}{\emph{{ACM} Comput. Surv.}} \bibinfo{volume}{51}, \bibinfo{number}{5} (\bibinfo{year}{2019}), \bibinfo{pages}{106:1--106:35}.
\newblock
\urldef\tempurl%
\url{https://doi.org/10.1145/3241741}
\showDOI{\tempurl}


\bibitem[Thakur et~al\mbox{.}(2024)]%
        {DBLP:conf/naacl/ThakurNAWLC24}
\bibfield{author}{\bibinfo{person}{Nandan Thakur}, \bibinfo{person}{Jianmo Ni}, \bibinfo{person}{Gustavo~Hern{\'{a}}ndez {\'{A}}brego}, \bibinfo{person}{John Wieting}, \bibinfo{person}{Jimmy Lin}, {and} \bibinfo{person}{Daniel Cer}.} \bibinfo{year}{2024}\natexlab{}.
\newblock \showarticletitle{Leveraging LLMs for Synthesizing Training Data Across Many Languages in Multilingual Dense Retrieval}. In \bibinfo{booktitle}{\emph{Proceedings of the 2024 Conference of the North American Chapter of the Association for Computational Linguistics: Human Language Technologies (Volume 1: Long Papers), {NAACL} 2024, Mexico City, Mexico, June 16-21, 2024}}, \bibfield{editor}{\bibinfo{person}{Kevin Duh}, \bibinfo{person}{Helena G{\'{o}}mez{-}Adorno}, {and} \bibinfo{person}{Steven Bethard}} (Eds.). \bibinfo{publisher}{Association for Computational Linguistics}, \bibinfo{pages}{7699--7724}.
\newblock
\urldef\tempurl%
\url{https://doi.org/10.18653/V1/2024.NAACL-LONG.426}
\showDOI{\tempurl}


\bibitem[Wei et~al\mbox{.}(2023)]%
        {DBLP:journals/bioinformatics/WeiLILL23}
\bibfield{author}{\bibinfo{person}{Chih{-}Hsuan Wei}, \bibinfo{person}{Ling Luo}, \bibinfo{person}{Rezarta Islamaj}, \bibinfo{person}{Po{-}Ting Lai}, {and} \bibinfo{person}{Zhiyong Lu}.} \bibinfo{year}{2023}\natexlab{}.
\newblock \showarticletitle{GNorm2: an improved gene name recognition and normalization system}.
\newblock \bibinfo{journal}{\emph{Bioinform.}} \bibinfo{volume}{39}, \bibinfo{number}{10} (\bibinfo{year}{2023}).
\newblock
\urldef\tempurl%
\url{https://doi.org/10.1093/BIOINFORMATICS/BTAD599}
\showDOI{\tempurl}


\bibitem[Wei et~al\mbox{.}(2016)]%
        {CDR:DBLP:journals/biodb/WeiPLDMLWL16}
\bibfield{author}{\bibinfo{person}{Chih{-}Hsuan Wei}, \bibinfo{person}{Yifan Peng}, \bibinfo{person}{Robert Leaman}, \bibinfo{person}{Allan~Peter Davis}, \bibinfo{person}{Carolyn~J. Mattingly}, \bibinfo{person}{Jiao Li}, \bibinfo{person}{Thomas~C. Wiegers}, {and} \bibinfo{person}{Zhiyong Lu}.} \bibinfo{year}{2016}\natexlab{}.
\newblock \showarticletitle{Assessing the state of the art in biomedical relation extraction: overview of the BioCreative {V} chemical-disease relation {(CDR)} task}.
\newblock \bibinfo{journal}{\emph{Database J. Biol. Databases Curation}}  \bibinfo{volume}{2016} (\bibinfo{year}{2016}).
\newblock
\urldef\tempurl%
\url{https://doi.org/10.1093/DATABASE/BAW032}
\showDOI{\tempurl}


\bibitem[Wei et~al\mbox{.}(2024)]%
        {10.1093/nar/gkae235}
\bibfield{author}{\bibinfo{person}{Chih-Hsuan Wei}, \bibinfo{person}{Alexis Allot}, \bibinfo{person}{Po-Ting Lai}, \bibinfo{person}{Robert Leaman}, \bibinfo{person}{Shubo Tian}, \bibinfo{person}{Ling Luo}, \bibinfo{person}{Qiao Jin}, \bibinfo{person}{Zhizheng Wang}, \bibinfo{person}{Qingyu Chen}, {and} \bibinfo{person}{Zhiyong Lu}.} \bibinfo{year}{2024}\natexlab{}.
\newblock \showarticletitle{{PubTator 3.0: an AI-powered literature resource for unlocking biomedical knowledge}}.
\newblock \bibinfo{journal}{\emph{Nucleic Acids Research}} \bibinfo{volume}{52}, \bibinfo{number}{W1} (\bibinfo{date}{04} \bibinfo{year}{2024}), \bibinfo{pages}{W540--W546}.
\newblock
\showISSN{0305-1048}
\urldef\tempurl%
\url{https://doi.org/10.1093/nar/gkae235}
\showDOI{\tempurl}
\showeprint{https://academic.oup.com/nar/article-pdf/52/W1/W540/58436124/gkae235.pdf}


\bibitem[Wei et~al\mbox{.}(2015)]%
        {gnormplus}
\bibfield{author}{\bibinfo{person}{Chih-Hsuan Wei}, \bibinfo{person}{Hung-Yu Kao}, {and} \bibinfo{person}{Zhiyong Lu}.} \bibinfo{year}{2015}\natexlab{}.
\newblock \showarticletitle{GNormPlus: An Integrative Approach for Tagging Genes, Gene Families, and Protein Domains}.
\newblock \bibinfo{journal}{\emph{BioMed Research International}} \bibinfo{volume}{2015}, \bibinfo{number}{1} (\bibinfo{year}{2015}), \bibinfo{pages}{918710}.
\newblock
\urldef\tempurl%
\url{https://doi.org/10.1155/2015/918710}
\showDOI{\tempurl}


\bibitem[Weikum et~al\mbox{.}(2021)]%
        {DBLP:journals/ftdb/WeikumDRS21}
\bibfield{author}{\bibinfo{person}{Gerhard Weikum}, \bibinfo{person}{Xin~Luna Dong}, \bibinfo{person}{Simon Razniewski}, {and} \bibinfo{person}{Fabian~M. Suchanek}.} \bibinfo{year}{2021}\natexlab{}.
\newblock \showarticletitle{Machine Knowledge: Creation and Curation of Comprehensive Knowledge Bases}.
\newblock \bibinfo{journal}{\emph{Found. Trends Databases}} \bibinfo{volume}{10}, \bibinfo{number}{2-4} (\bibinfo{year}{2021}), \bibinfo{pages}{108--490}.
\newblock
\urldef\tempurl%
\url{https://doi.org/10.1561/1900000064}
\showDOI{\tempurl}


\bibitem[Wishart et~al\mbox{.}(2006)]%
        {Drugbank:DBLP:journals/nar/WishartKGSHSCW06}
\bibfield{author}{\bibinfo{person}{David~S. Wishart}, \bibinfo{person}{Craig Knox}, \bibinfo{person}{Anchi Guo}, \bibinfo{person}{Savita Shrivastava}, \bibinfo{person}{Murtaza Hassanali}, \bibinfo{person}{Paul Stothard}, \bibinfo{person}{Zhan Chang}, {and} \bibinfo{person}{Jennifer Woolsey}.} \bibinfo{year}{2006}\natexlab{}.
\newblock \showarticletitle{DrugBank: a comprehensive resource for \emph{in silico} drug discovery and exploration}.
\newblock \bibinfo{journal}{\emph{Nucleic Acids Res.}} \bibinfo{volume}{34}, \bibinfo{number}{Database-Issue} (\bibinfo{year}{2006}), \bibinfo{pages}{668--672}.
\newblock
\urldef\tempurl%
\url{https://doi.org/10.1093/NAR/GKJ067}
\showDOI{\tempurl}


\bibitem[Yan and Chen(2022)]%
        {DBLP:conf/eeke/YanC22}
\bibfield{author}{\bibinfo{person}{Yuchen Yan} {and} \bibinfo{person}{Chong Chen}.} \bibinfo{year}{2022}\natexlab{}.
\newblock \showarticletitle{SciGraph: {A} Knowledge Graph Constructed by Function and Topic Annotation of Scientific Papers (poster)}. In \bibinfo{booktitle}{\emph{3rd Workshop on Extraction and Evaluation of Knowledge Entities from Scientific Documents, EEKE@JCDL 2022, Germany and Online, 23-24 June, 2022}} \emph{(\bibinfo{series}{{CEUR} Workshop Proceedings}, Vol.~\bibinfo{volume}{3210})}. \bibinfo{publisher}{CEUR-WS.org}, \bibinfo{pages}{134--137}.
\newblock
\urldef\tempurl%
\url{https://ceur-ws.org/Vol-3210/paper16.pdf}
\showURL{%
\tempurl}


\bibitem[Yang et~al\mbox{.}(2019)]%
        {DBLP:conf/nips/YangDYCSL19}
\bibfield{author}{\bibinfo{person}{Zhilin Yang}, \bibinfo{person}{Zihang Dai}, \bibinfo{person}{Yiming Yang}, \bibinfo{person}{Jaime~G. Carbonell}, \bibinfo{person}{Ruslan Salakhutdinov}, {and} \bibinfo{person}{Quoc~V. Le}.} \bibinfo{year}{2019}\natexlab{}.
\newblock \showarticletitle{XLNet: Generalized Autoregressive Pretraining for Language Understanding}. In \bibinfo{booktitle}{\emph{Advances in Neural Information Processing Systems 32: Annual Conference on Neural Information Processing Systems 2019, NeurIPS 2019, December 8-14, 2019, Vancouver, BC, Canada}}. \bibinfo{pages}{5754--5764}.
\newblock
\urldef\tempurl%
\url{https://proceedings.neurips.cc/paper/2019/hash/dc6a7e655d7e5840e66733e9ee67cc69-Abstract.html}
\showURL{%
\tempurl}


\bibitem[Yao et~al\mbox{.}(2019)]%
        {DBLP:conf/aaai/YaoM019}
\bibfield{author}{\bibinfo{person}{Liang Yao}, \bibinfo{person}{Chengsheng Mao}, {and} \bibinfo{person}{Yuan Luo}.} \bibinfo{year}{2019}\natexlab{}.
\newblock \showarticletitle{Graph Convolutional Networks for Text Classification}. In \bibinfo{booktitle}{\emph{The Thirty-Third {AAAI} Conference on Artificial Intelligence, {AAAI} 2019, The Thirty-First Innovative Applications of Artificial Intelligence Conference, {IAAI} 2019, The Ninth {AAAI} Symposium on Educational Advances in Artificial Intelligence, {EAAI} 2019, Honolulu, Hawaii, USA, January 27 - February 1, 2019}}. \bibinfo{publisher}{{AAAI} Press}, \bibinfo{pages}{7370--7377}.
\newblock
\urldef\tempurl%
\url{https://doi.org/10.1609/AAAI.V33I01.33017370}
\showDOI{\tempurl}


\bibitem[Yasunaga et~al\mbox{.}(2022)]%
        {DBLP:conf/acl/YasunagaLL22}
\bibfield{author}{\bibinfo{person}{Michihiro Yasunaga}, \bibinfo{person}{Jure Leskovec}, {and} \bibinfo{person}{Percy Liang}.} \bibinfo{year}{2022}\natexlab{}.
\newblock \showarticletitle{LinkBERT: Pretraining Language Models with Document Links}. In \bibinfo{booktitle}{\emph{Proceedings of the 60th Annual Meeting of the Association for Computational Linguistics (Volume 1: Long Papers), {ACL} 2022, Dublin, Ireland, May 22-27, 2022}}. \bibinfo{publisher}{Association for Computational Linguistics}, \bibinfo{pages}{8003--8016}.
\newblock
\urldef\tempurl%
\url{https://doi.org/10.18653/V1/2022.ACL-LONG.551}
\showDOI{\tempurl}


\end{thebibliography}

\end{document}